\documentclass[11pt,onecolumn,a4paper]{article}

\usepackage{amsmath}
\usepackage{amssymb}
\usepackage{graphicx}
\usepackage{bm}
\usepackage{tikz}
\usetikzlibrary{automata, arrows, positioning, calc, bending, decorations.text, decorations.pathreplacing, angles, quotes, fit, patterns}
\usepackage[utf8]{inputenc}
\usepackage{chemformula}
\usepackage{epsfig}
\usepackage{url}

\usepackage{enumitem}
\usepackage{subfig}
\usepackage{epstopdf}
\usepackage{svg}
\usepackage{forest}
\usepackage{algorithm}
\usepackage[noend]{algpseudocode}
\usepackage{booktabs}
\usepackage{pgfplots}
\usepackage{siunitx}
\usetikzlibrary{decorations.pathreplacing,decorations.markings}
\tikzset{
  font=\normalsize,
  red arrow/.style={
    midway,red,sloped,fill, minimum height=1.5cm, single arrow, single arrow head extend=.6cm, single arrow head indent=.25cm,xscale=0.3,yscale=0.15,
    allow upside down
  },
  black arrow/.style 2 args={-stealth, shorten >=#1, shorten <=#2},
  black arrow/.default={1mm}{1mm},
  tree box/.style={draw, rounded corners, inner sep=.3em},
  node box/.style={white, draw=black, text=black, rectangle, rounded corners},
}

\usepackage{xcolor}

\title{\Large{Minimal set of crystallographic descriptors for sorption properties in hypothetical Metal Organic Frameworks: Role in sequential learning optimization}}
\author{Giovanni Trezza, Luca Bergamasco, Matteo Fasano, Eliodoro Chiavazzo\thanks{Corresponding author: eliodoro.chiavazzo@polito.it}\\ \small{\emph{Department of Energy, Politecnico di Torino, C.so Duca degli Abruzzi 24, Torino 10129, Italy}}}
\date{}

\usepackage[left=2cm, right=2cm, top=2cm]{geometry}

\begin{document}

\maketitle

\begin{abstract}
%
%
%
%
Several studies have been recently reported in the literature on sorption properties of MOFs with a number of organic sorbates, such as ethanol and methanol.
Surprisingly, still few studies have been reported on water sorbate despite its large availability, low cost and environmental sustainability, and the screening of a large number of hypothetical MOFs-water working pairs for engineering applications is still challenging.
%
%
Based on a recently reported database of over 5000 hypothetical MOFs, a first contribution of this study is the identification of the minimal set of crystallographic descriptors underpinning the most important sorption properties of MOFs for \ch{CO2} and, importantly, for \ch{H2O}.
%
Furthermore, a comprehensive comparison of several Sequential Learning (SL) algorithms for MOFs properties optimization is carried out and the role played by the above minimal set of crystallographic descriptors clarified.
%
In sorption-based energy transformations, thermodynamic limits of important figures of merit (e.g. maximum specific energy) depend both on operating conditions and equilibrium sorption properties in a wide range of sorbate coverage values.
The access to the latter properties is often incomplete, with essential quantities such as equilibrium adsorption isotherms spanning over the full sorbate coverage range and values of the isosteric heat being only partially available.
As a result, this may prevent the computation of objective functions during the optimization procedure.
We propose a fast procedure for optimizing specific energy in a closed sorption energy storage system with the only access to the water Henry coefficient at a fixed temperature value and to the specific surface area.
%
%
\begin{description}
\item[Keywords] \emph{Metal Organic Frameworks, Sequential Learning, Thermal Energy Storage}
\end{description}
\end{abstract}

\section*{Introduction\label{sec:level1}}

Metal Organic Frameworks (MOFs) are crystalline compounds consisting of metal ions and organic linkers, characterized by tunable porosity and incredibly high surface area \cite{kitagawa2014metal}. 
Due to their unique properties, MOFs have recently attracted remarkable attention in a wide range of different fields, including gas/vapour separation \cite{adil2017gas}, reaction catalysis \cite{rogge2017metal}, drug delivery \cite{wuttke2017positioning}, energy storage and heat transformations \cite{chaemchuen2018tunable, de2015adsorption}. 
Given their nature of porous adsorbent materials, an intensively active research is focused on the use of MOFs for \ch{CO2} capture, towards the development of effective technologies for mitigating green-house gas emissions \cite{chen2016understanding}.
Recently, MOFs have been also employed in adsorption-based atmospheric water harvesting driven by solar thermal energy \cite{kim2018adsorption,kalmutzki2018metal}. 
In general, when dealing with application of engineering relevance, different inlet gas streams, variable operating conditions, and target properties tailored per each specific case make it challenging to identify an ideal MOF crystal for all applications \cite{ejeian2021adsorption}, thus leading to a fragmented case-by-case optimization problem.

Hence, MOFs require proper and efficient methods for tailoring their features according to target properties of interest in each specific application.
The latter is everything but an easy task. 
In fact, due to the myriad of degrees of freedom for MOFs structure and composition, more than 100 trillion compounds have been hypothesized \cite{lee2021computational}, while almost 100000 have been synthesized so far \cite{li2020enabling}. 
High-throughput computational screening and machine learning have been recently adopted to analyse large MOF datasets.
Such computational tools allow to identify significant correlations between nanoscale features and observable macroscale properties \cite{anderson2018role,moghadam2019structure}, and to select the most suitable crystal for a given application case.
A few representative examples are provided by gas-gas separation (\ch{D2/H2} \cite{zhou2020toward}, \ch{O2/N2} \cite{yan2022machine}, \ch{CO/N2} \cite{rampal2021development}, \ch{CO2/H2} \cite{avci2020new}, ethane/ethylene \cite{halder2020high}, and other gas mixtures \cite{yang2019computational}), enantioselectivity of chemical compounds \cite{qiao2021molecular}, gas adsorption (\ch{CO2} \cite{li2016high}, \ch{CH4} \cite{pardakhti2017machine}, \ch{H2} \cite{bobbitt2019molecular}, thiol \cite{qiao2017high}, organosulfurs \cite{liang2019combining}, acetylene \cite{yang2021analyzing}) and combinations thereof \cite{dureckova2019robust,liu2021predicting}. 
Several computational explorations of MOFs datasets have been carried out also for biomedical (drug delivery \cite{ma2020computer}), mechanical (\ch{CO2} Brayton cycle \cite{du2021high}, osmotic heat engine \cite{long2021screening}) and energy applications (heat pumps/chillers \cite{shi2020machine,shi2021techno}, thermal energy storage \cite{garcia2021systematic}).

In this context, modern \emph{Sequential Learning} (SL) algorithms are emerging as particularly efficient tools for exploring the material high dimensional (crystallographic) feature space.
In particular, while evaluating an objective black-box function through demanding physical or numerical experiments, SL tools can provide a well-orchestrated procedure to rationally navigate the high-dimensional parameter (feature) space.
Thus, given an initial pool of evaluation points, one can sequentially choose the next experiment to carry out \cite{brochu2010tutorial, ahmadi2013predicting}, without relying on a naive random guessing.
%
%
Rather general techniques have been proposed in the area of material science holding the promise to accelerate materials discovery and research \cite{rohr2020benchmarking}, with a number of Authors reporting successful use of SL approaches in this field. 
Aggarwal \emph{et al.} \cite{aggarwal2016information}, by means of optimal experimental design, successfully characterized a substrate under a thin film. 
Seko \emph{et al.} \cite{seko2014machine} found the compound with the highest melting temperature in a given ensemble of candidate materials with less attempts than a naive random choice. 
Kiyohara \emph{et al.} \cite{kiyohara2016acceleration} accelerated the search of a stable interface structure with respect to a traditional \emph{brute force} approach. 
Dehghannasiri \emph{et al.} \cite{dehghannasiri2017optimal} efficiently guided experiments to design the shape memory alloy with the lowest energy dissipation at a given temperature. 
%
Needless to say that the identification of the parameter space is a very important preliminary step when implementing SL algorithms.
%
%
Here, we choose to specifically focus on MOFs properties as gas/vapour sorbent materials, since those are particularly relevant for energy applications. 

The first important objective of this work is the identification of the minimal set of MOFs features (or {\it descriptors}) ruling critical adsorption properties in the low-coverage regime, i.e. the Henry solubility coefficients for both \ch{CO2}-MOFs and, importantly, \ch{H2O}-MOFs working pairs.
The above minimal set represent the important crystallographic features underpinning a given adsorption property of interest. 
In this sense, each minimal set of descriptors is here referred to as the {\it genetic code} for a given property, and it is identified as described below.

First, we curate and enhance MOF data from a recently developed library made of 8206 compounds generated computationally \cite{boyd2019data}. 
Each Crystallographic Information File (CIF) representing a given material is first featurized by means of 1557 Classical Force-field Inspired Descriptors (CFID) \cite{choudhary2018machine} taking into account both chemical and structural parameters.
Subsequently, we train and validate regression models of target properties involved in heat storage applications, such as Henry coefficients and working capacity. 
These models are obtained by means of AutoMatminer \cite{dunn2020benchmarking}, which is able to automatically find the best pipeline, clean the database, reduce the features, choose the model and tune its hyperparameters. 
The final ranking and selection of the minimal set of descriptors is finally performed by evaluating the importance of each feature on the models outputs by means of the Kernel SHAP interpretation algorithm \cite{NIPS2017_7062}.

\begin{figure*}[ht]
\includegraphics[width=1.0\textwidth]{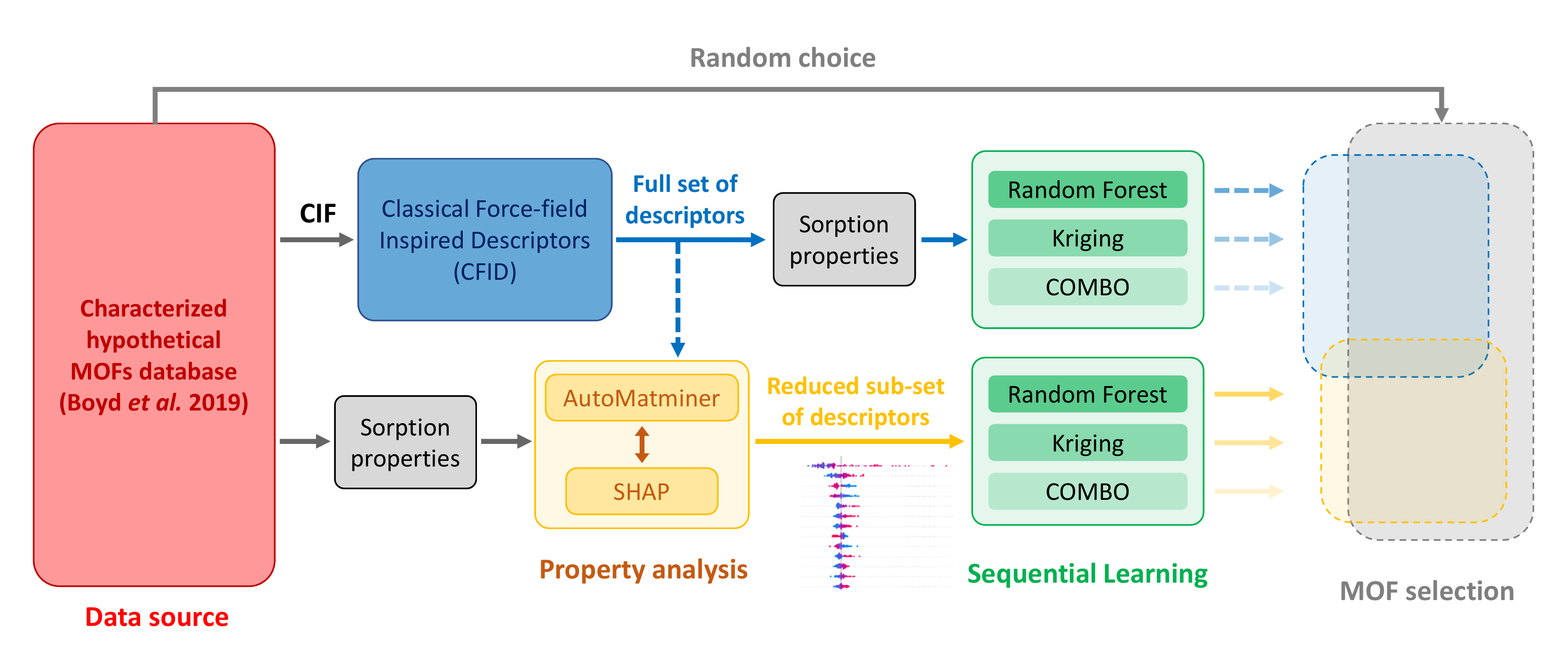}
\caption{Overview of the methodological protocols to identify and test the minimal set of ruling crystallographic descriptors of sorption properties in several sequential learning algorithms. Over 5000 hypothetical MOFs from ref. \cite{boyd2019data} are first featurized by CFID, with the corresponding full set of descriptors provided to AutoMatminer for a preliminary descriptor reduction and machine learning models training of sorption properties of interest. The Kernel SHAP interpretation algorithm is thus used to finalize the identification and ranking of a reduced sub-set of ruling descriptors ({\it genetic code} of the chosen property). Several sequential learning schemes are tested using both the full set of descriptors and the reduced one for comprehensive comparison.}
\label{fig:methodology}
\end{figure*}

Upon identification of the above crystallographic {\it genetic code} of sorption properties in MOFs, we investigate its role when using SL algorithms. 
Therefore, we compare the performance of three different SL methodologies aiming at maximizing \ch{H2O} and \ch{CO2} Henry coefficients, and \ch{CO2} working capacity: (a) random Forests with Uncertainty Estimates for Learning Sequentially (FUELS \cite{ling2017high}); (b) kriging algorithm \cite{lophaven2002matlab}; (c) COMmon Bayesian Optimization Library (COMBO) \cite{ueno2016combo}. 
For each SL methodology, we compare several strategies for choosing the next material to test, combining the \emph{exploration} of high-uncertainty regions with the \emph{exploitation} of high-performing candidates. 
Importantly, we analyse the SL performance using both the minimal subset of features (from AutoMatminer and SHAP analysis) and a larger set of variables, to highlight how the identification of descriptors affects the minimum number of experiments needed to pick out a MOF with the highest value of a desired property. 
In Fig. \ref{fig:methodology} the above procedure is schematically represented. 

We highlight that sorption-based engineering applications rely upon sorbent material characterization in a wide coverage range. 
However, when a large number of hypothetical sorbents (here MOFs, but also zeolites in principle \cite{fasano2019water,fasano2019mechanistic}) have to be evaluated as potential candidates, only low-coverage characterization (i.e. Henry coefficient) is often accessible thus making challenging any optimization of crucial figures of merits of engineering relevance. 
We thus formulate an innovative procedure aiming at a fast evaluation of one of the most important figures of merit in closed water sorption seasonal thermal energy storage applications, namely the material-based specific (stored) energy.
Unlike traditional sensible or latent systems \cite{neri2020numerical}, the above sorption based energy storage technologies have the unique advantage to be loss-free.
Our procedure can thus be used in SL-based (or other) optimization/screening processes of MOFs even under incomplete knowledge of the entire isosteric field of the candidate working pairs.
Applied to the database of over 5000 computationally generated (hypothetical) compounds by ref. \cite{boyd2019data} (developed for different purposes), our procedure identifies MOFs capable to largely outperform state-of-the-art sorbent materials for thermal energy storage.

\section*{Results}
%
%
%
In this work, a crucial source of data on MOFs stems from the dataset of Boyd {\it et al.} \cite{boyd2019data}, where important sorption properties (e.g. the Henry coefficients for \ch{CO_2} and \ch{H_2O}, the working capacity for \ch{CO_2}, the specific surface area) have been computed by DFT-based simulations for over 8000 potential MOFs.
Capitalizing on the above comprehensive study, we construct Machine Learning (ML) models capable of accurately predicting MOFs solubility of both \ch{CO_2} and \ch{H2O} as well as \ch{CO_2} working capacity and surface area.
The above models enable us to achieve the first important result of this work, namely the identification of the minimal set of crystallographic-based descriptors \cite{ward2018matminer} ruling these sorption and geometric properties in MOFs.

Moreover, a systematic comparison of SL approaches on the above MOFs database reveals important conclusions on the performance of the different regression schemes adopted and, most importantly, the role played by the selection of the feature space to be explored.
Those conclusions are also supported by results obtained on a highly-controllable synthetic dataset, as discussed in detail in the Supplementary Section S1.

It is worth stressing that properties reported in ref. \cite{boyd2019data} only characterize MOFs in the Henry regime and are not sufficient to describe the equilibrium sorption properties in the high coverage regime.
However, when targeting important engineering applications such as seasonal thermal energy storage, key figures of merits of the storage plant (e.g. the material-based specific energy) critically rely upon the access to the entire isosteric field of the chosen sorbent-sorbate pair or, equivalently, to the knowledge of equilibrium adsorption isotherms at several temperature values \cite{fasano2016,fasano2019atomistic}.
The latter isotherms describe the adsorption properties (at equilibrium) of sorbents in a wide range of coverage values, from the Henry regime up to the saturation pressure of the sorbate fluid.

Therefore, in this work, we also propose an innovative approach enabling us to optimize one of the most important engineering figures of merit of MOFs for seasonal thermal energy storage applications (i.e. material-based specific energy in an ideal closed sorption cycle, see Methods), even if only an incomplete set of sorption properties are (experimentally or numerically) accessible.
Based on the latter optimization procedure, we are finally able to identify potential MOFs candidates for seasonal thermal energy storage that largely outperform most of the current state-of-the-art sorbent materials.
%
%
%
%
%

\subsection*{Descriptors of sorption properties in MOFs and their use in SL algorithms}

We constructed four MOFs datasets, each one with the same 1557 features and a different target property among Henry coefficient for \ch{CO2} (8194 data entries), working capacity for \ch{CO2} (8202 data entries), Henry coefficient for \ch{H2O} (8202 data entries) and surface area (5028 data entries). The different number of data entries are due to missing values for some of the chosen properties in the available database by ref. \cite{boyd2019data}.

The above database also reported, for all the compounds, both the crystallographic file (used to extract the 1557 Classical Force-Field Inspired Descriptors - CFID \cite{choudhary2018machine} by means of Matminer \cite{ward2018matminer}), and a list of DFT-computed properties, among which we have only considered the above mentioned adsorption properties of interest.
More specifically, the computed 1557 explanatory variables (also referred to as {\it descriptors}) proposed by Choudhary \emph{et al.} \cite{choudhary2018machine} consist of a set of both chemical (e.g., average chemical properties over the elements in the cell, average atomic radial charge) and structural (e.g. distribution functions) quantities.
More details on descriptor sub-categories are reported in Table~\ref{tab1}.

\begin{table*}[h]
\centering
\caption{\label{tab1}Components of Classical Force-Field Inspired Descriptors (CFID) \cite{choudhary2018machine}.}
\begin{tabular}{lr}
\toprule
\textbf{Descriptor name} & \textbf{Total number}\\
\midrule
Chemical & 438\\
Simulation cell-size & 4\\
Radial charge & 378\\
Radial distribution function & 100\\
Angular distribution up to first nearest neighbor cutoff & 179\\
Angular distribution up to second nearest neighbor cutoff & 179\\
Dihedral distribution up to first nearest neighbor cutoff & 179\\
Nearest neighbor distribution & 100\\
\hline

\textbf{Total} & \textbf{1557}\\
\bottomrule
\end{tabular}
\end{table*}

%
Similarly to the synthetic case in Supplementary Section S1, we have trained four different ML models by means of AutoMatminer to predict the Henry coefficient for \ch{CO2}, the working capacity for \ch{CO2}, the Henry coefficient for \ch{H2O} and the surface area, achieving coefficients of determination of $R^2=0.803$, $R^2=0.655$, $R^2=0.885$ and $R^2=0.925$, respectively. 
We have used 80\% of each dataset to train the models, and the remaining 20\% to validate them.
Since the Henry coefficient values span a few orders of magnitude, the corresponding ML models have been developed in terms of the natural logarithm of those properties.
During the data pre-processing routines, each of the three AutoMatminer pipelines (i.e., feature reduction, data cleaning and machine learning with automatic hyper-parameter tuning, see Methods and Supplementary Section S3 for details) already drops a significant number of the 1557 features, thus confirming that many of the initially selected descriptors do not significantly affect the chosen adsorption properties.
More specifically, the final models include 44 descriptors for \ch{CO2} Henry coefficient, 92 descriptors for \ch{CO2} working capacity, 36 descriptors for \ch{H2O} Henry coefficient and 24 descriptors for surface area. 

Then, the Kernel SHAP routine \cite{NIPS2017_7062} allows to identify the most meaningful descriptors as those accounting for the 90\% of the cumulative curve over the coefficients of importance. 
The SHAP routine identifies the most meaningful features for the trained models among the complete set of 1557 CFID, ending up with a subset of those retained by AutoMatminer after the pre-processing.
In particular, the impact of a descriptor depends on the comparison between the output of a model trained with that feature and another model output, trained without that feature (see Methods). 
The coefficients of importance are thus computed over the testing set, i.e., over samples the model has never encountered during the training. 
Since the agnostic realization of the Kernel SHAP routine is memory demanding, we have taken into account only 100 random samples from the testing set to evaluate the impact of features.

\begin{figure*}
\centering
\includegraphics[width=.34\textwidth]{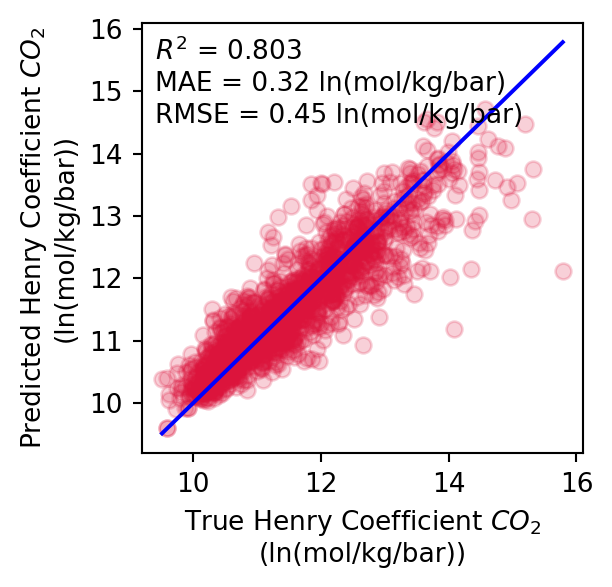}\quad 
\includegraphics[width=.34\textwidth]{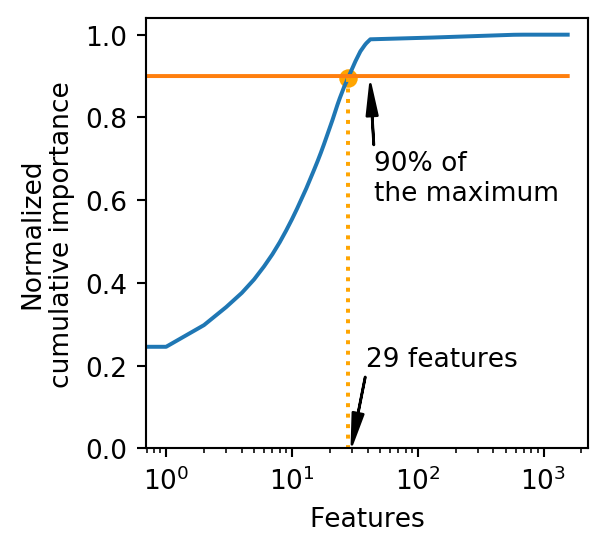} 
\includegraphics[width=.34\textwidth]{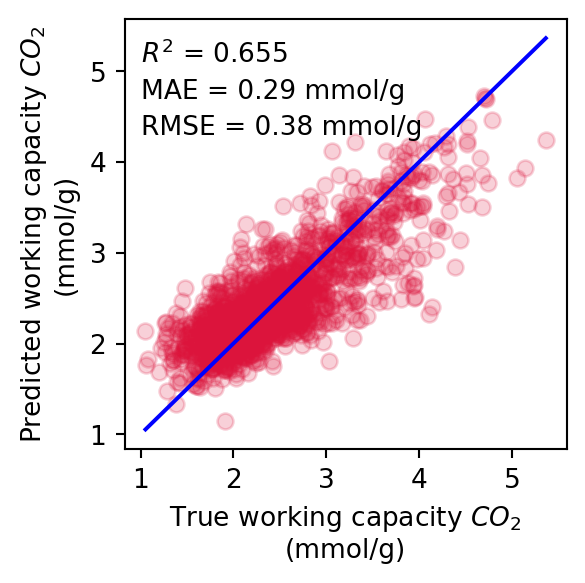}\quad
\includegraphics[width=.34\textwidth]{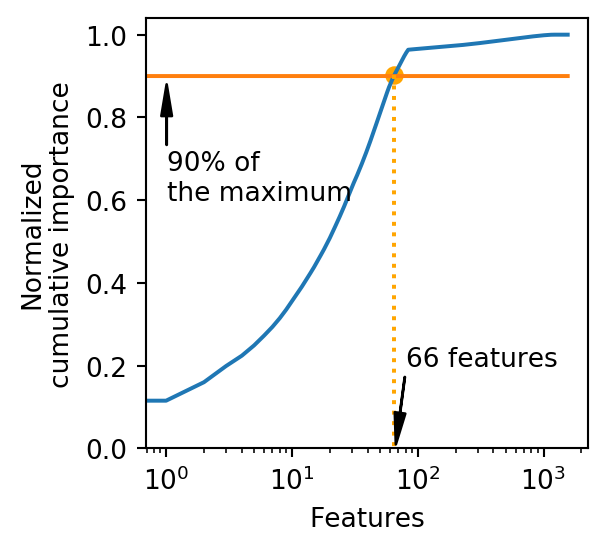}
\includegraphics[width=.34\textwidth]{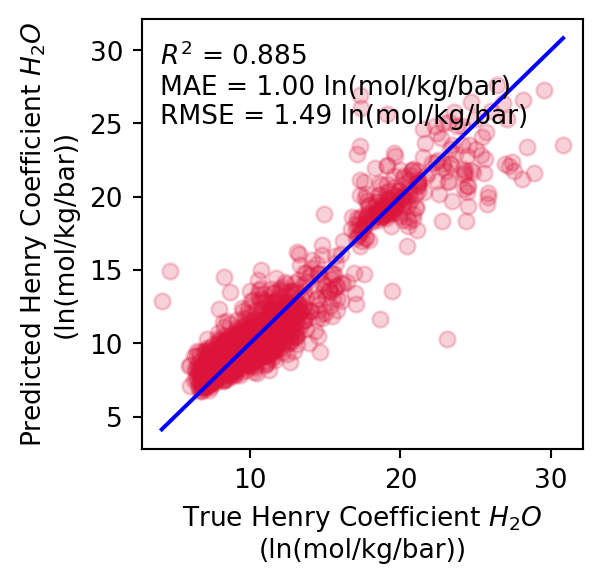} \quad
\includegraphics[width=.34\textwidth]{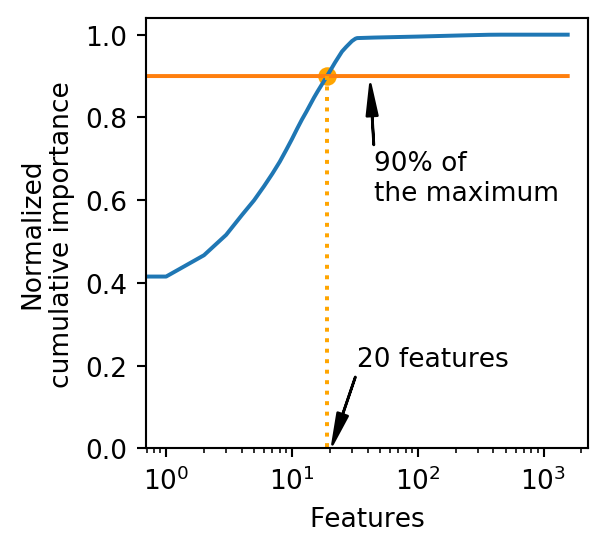} 
\includegraphics[width=.34\textwidth]{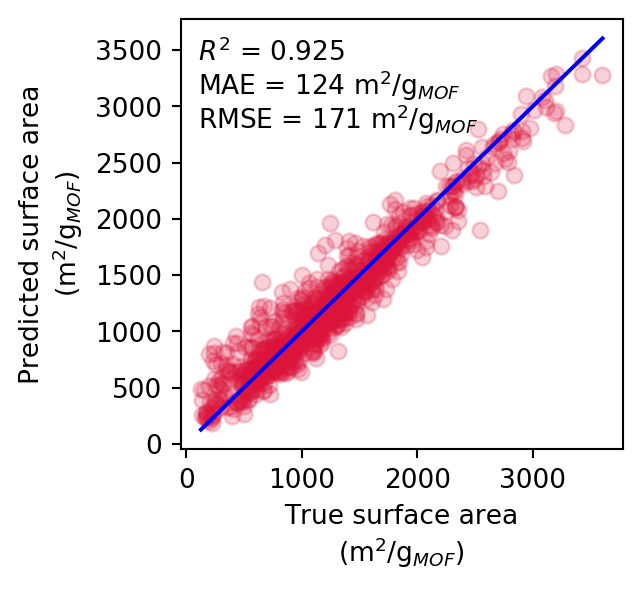}\quad
\includegraphics[width=.34\textwidth]{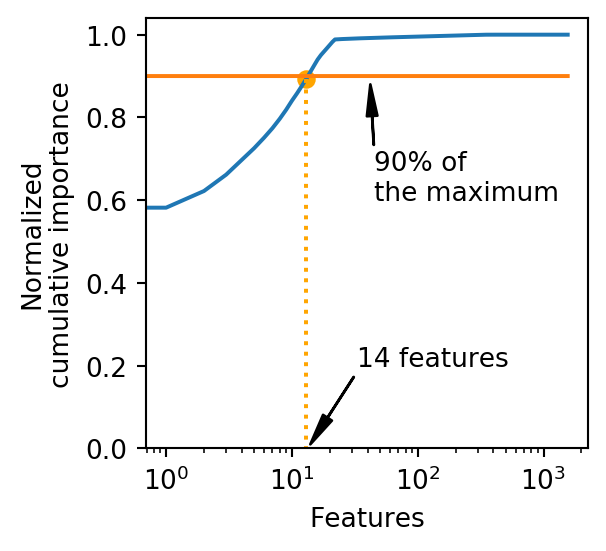}

\caption{Predictions and corresponding normalized cumulative curves for the coefficients of importance of the four regression AutoMatminer models trained for Henry coefficient for \ch{CO2}, working capacity for \ch{CO2}, Henry coefficient for \ch{H2O}, surface area. Model performance is shown in terms of coefficient of determination $R^2$, Mean Absolute Error (MAE), Root Mean Squared Error (RMSE).}
\label{fig:models}
\end{figure*}

\begin{figure}
\centering
\includegraphics[width = 0.7\textwidth]{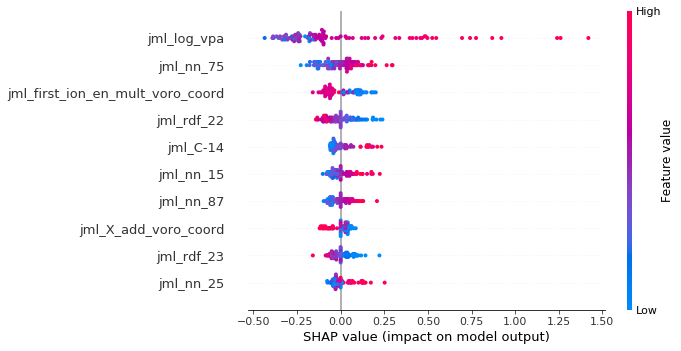}
\caption{10 most important features according to SHAP ranking of Henry coefficient for \ch{CO2}. For each feature (i.e., each line), 100 dots are shown, representing the 100 samples of the testing set used for computing 100 different SHAP values (impacts on the model output, horizontal axis). The color represents the corresponding feature value. The features are sorted according to the mean over the absolute SHAP values.}
\label{fig:SHAPco2}
\end{figure}

\begin{figure}
\centering
\includegraphics[width = 0.7\textwidth]{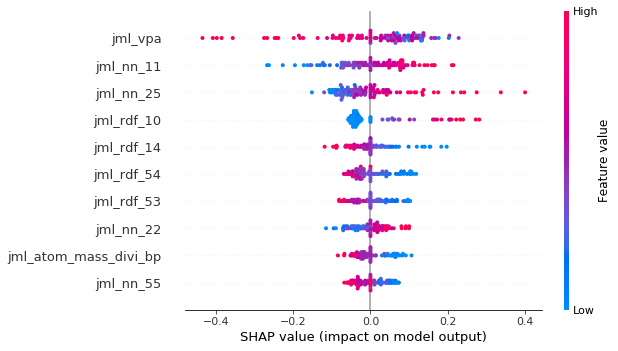}
\caption{10 most important features according to SHAP ranking of working capacity for \ch{CO2}. For each feature (i.e., each line), 100 dots are shown, representing the 100 samples of the testing set used for computing 100 different SHAP values (impacts on the model output, horizontal axis). The color represents the corresponding feature value. The features are sorted according to the mean over the absolute SHAP values.}
\label{fig:SHAPworking_capacity}
\end{figure}

\begin{figure}
\centering
\includegraphics[width = 0.7\textwidth]{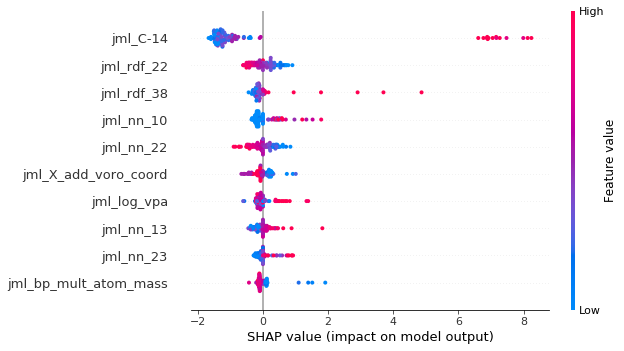}
\caption{10 most important features according to SHAP ranking of Henry coefficient for \ch{H2O}. For each feature (i.e., each line), 100 dots are shown, representing the 100 samples of the testing set used for computing 100 different SHAP values (impacts on the model output, horizontal axis). The color represents the corresponding feature value. The features are sorted according to the mean over the absolute SHAP values.}
\label{fig:SHAPh2o}
\end{figure}

\begin{figure}
\centering
\includegraphics[width = 0.7\textwidth]{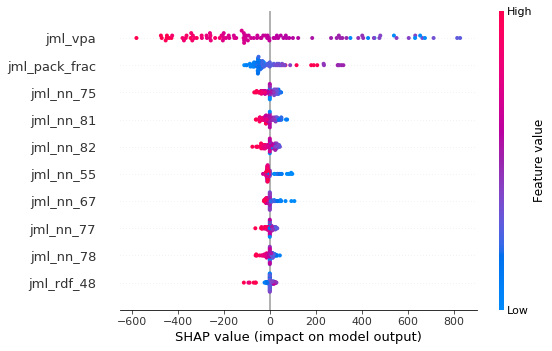}
\caption{10 most important features according to SHAP ranking of surface area. For each feature (i.e., each line), 100 dots are shown, representing the 100 samples of the testing set used for computing 100 different SHAP values (impacts on the model output, horizontal axis). The color represents the corresponding feature value. The features are sorted according to the mean over the absolute SHAP values.}
\label{fig:SHAPsurface}
\end{figure}

Overall, starting from the original 1557 Classical Force field Inspired Descriptors, 29 items for the \ch{CO2} Henry coefficient, 66 for the \ch{CO2} working capacity, 20 for the \ch{H2O} Henry coefficient and 14 for surface area are found to explain 90\% of the corresponding regression models.
These minimal sets of ruling descriptors are reported in Fig.~\ref{fig:models} with the corresponding cumulative importance curves, while Figs.~\ref{fig:SHAPco2},  \ref{fig:SHAPworking_capacity}, \ref{fig:SHAPh2o} and \ref{fig:SHAPsurface} show the SHAP rankings of the ten most meaningful descriptors. 
Table \ref{tab6} summarizes the physicochemical meaning of the identified descriptors, based on the complete list by Choudhary \emph{et al.} \cite{choudhary2018machine}. 
The list of the AutoMatminer retained variables are shown in the Supplementary Section S2, together with their cumulative importance percentage according to the SHAP rankings.
As far as the four properties of MOF are concerned, the identification of those important descriptors represents {\it per se} an advancement of knowledge on sorption mechanism in MOFs and a first important contribution of this work.

\begin{table*}
\caption{\label{tab6} Physicochemical meaning of CFID descriptors \cite{choudhary2018machine}.}
\begin{tabular}{ll}
\toprule
\textbf{CFID descriptors} & \textbf{Meaning}\\
\midrule
`jml\_jv-enp' & energy per atom of an element from JARVIS-DFT\\
`jml\_X' & electronegativity\\
`jml\_bp' & boiling point\\
`jml\_mol\_vol' & molar volume\\
`jml\_atom\_rad' & atomic radii\\
`jml\_atom\_mass' & atomic mass\\
`jml\_voro\_coord' & Voronoi coordination number of an elemental-crystal structure\\
`jml\_hfus' & heat of fusion of an element\\
`jml\_C-14' & elastic constant of the 14th element from JARVIS-DFT\\
`jml\_polzbl' & polarizability\\
`jml\_first\_ion\_en' & first ionization energy of an element\\
`jml\_elec\_aff' &  electron affinity\\
`jml\_vpa' & volume per atom of the cell\\
`jml\_log\_vpa' & logarithm of volume per atom of the cell\\
`jml\_pack\_frack' & packing fraction\\
`jml\_mean\_chg\_\#' & \#-th descriptor of radial charge\\
`jml\_rdf\_\#' & \#-th descriptor of radial distribution function\\
`jml\_adf1\_\#' & \#-th descriptor of angular distribution up to first nearest neighbor cutoff\\
`jml\_nn\_\#' & \#-th descriptor of nearest neighbor distribution\\
`add', `mult', `divi' & addition, multiplication, quotient between different descriptors\\
\bottomrule
\end{tabular}
\end{table*}

\begin{figure*}
\centering
\includegraphics[width=0.75\textwidth]{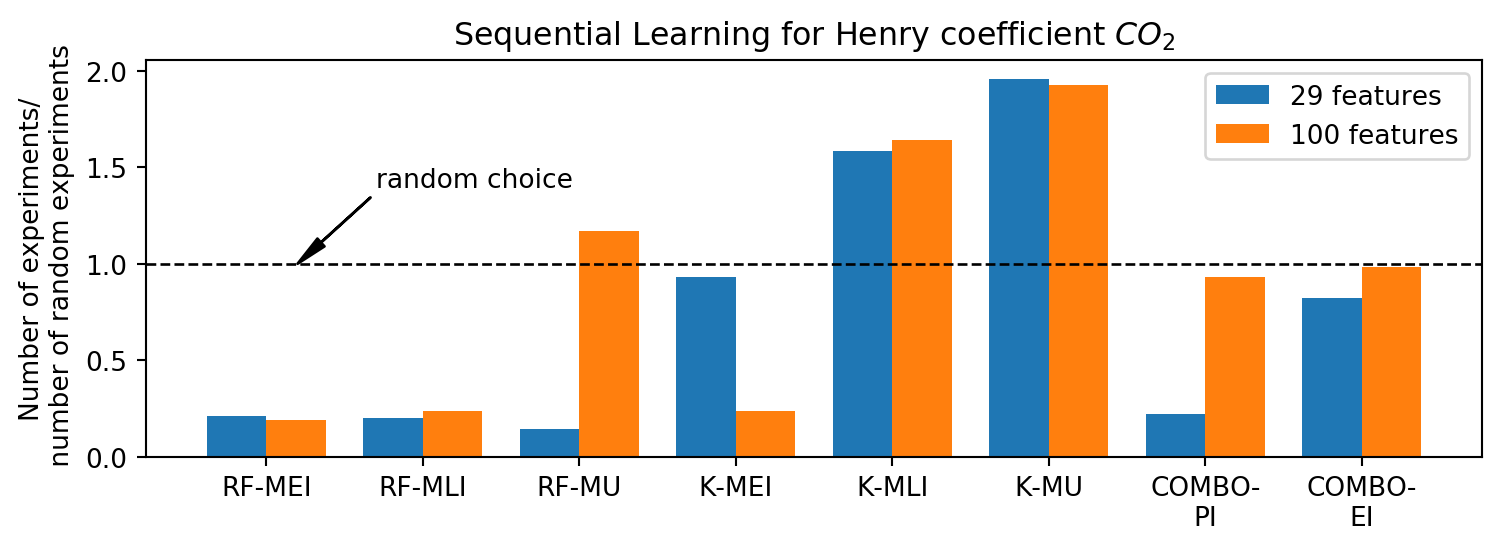}
\includegraphics[width=0.75\textwidth]{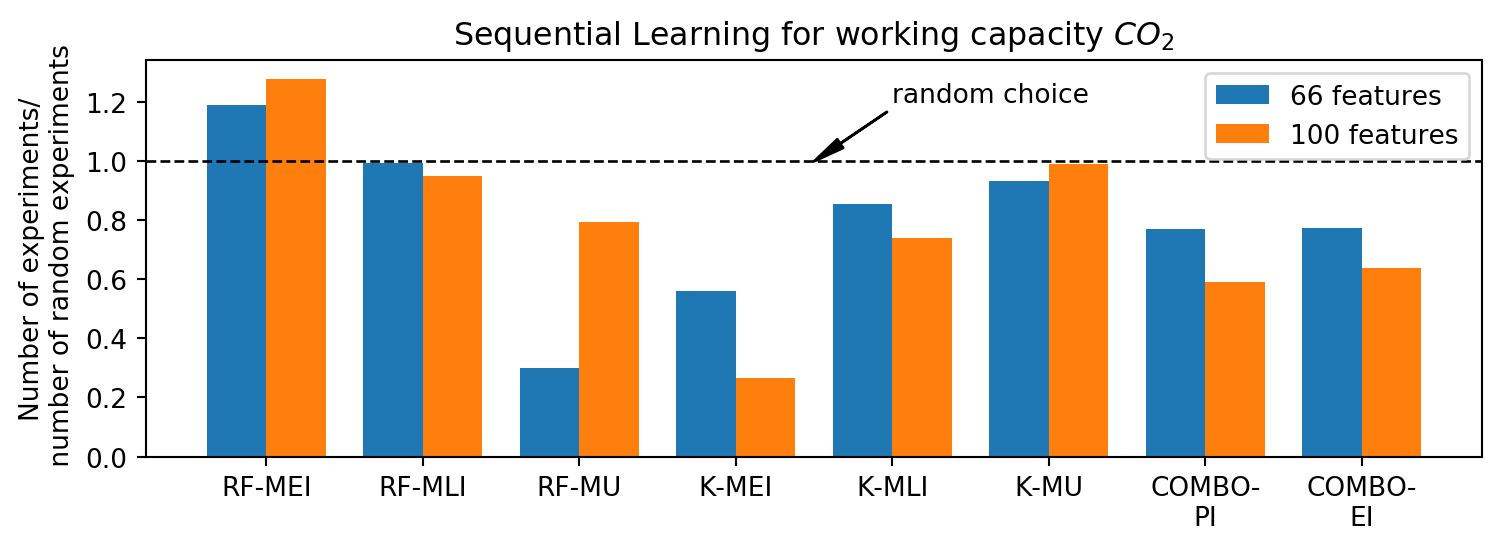}
\includegraphics[width=0.75\textwidth]{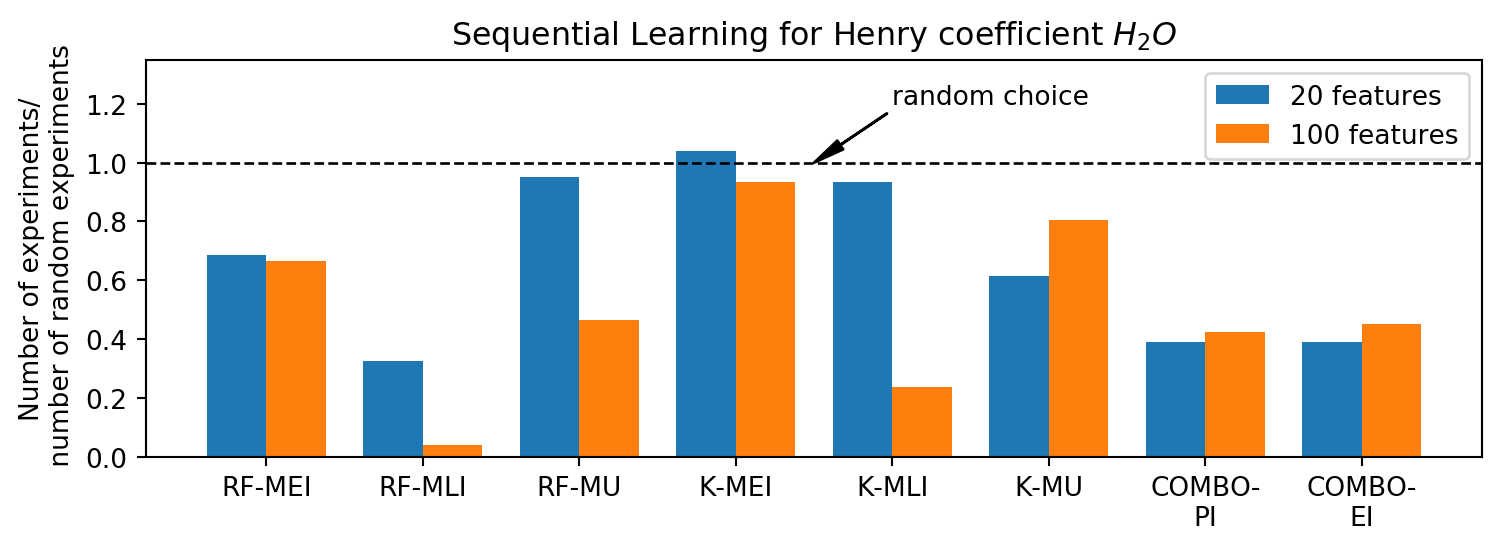}
\caption{Number of evaluations before converging to the maximum for the SL algorithms, normalized with respect to the random choice (corresponding to 200 experiments), for three sorption properties of MOFs: Henry coefficient for \ch{CO2}, working capacity for \ch{CO2}, Henry coefficient for \ch{H2O}. The initial set consists of the same worst 100 candidates (in terms of the target property) from a random subset of 500 samples of the original database.}
\label{fig:SL}
\end{figure*}

Upon the identification of the above lists of descriptors, we have compared the performance of SL algorithms for the sorption properties of interest using both the reduced set of important descriptors (i.e. those in Figs. \ref{fig:SHAPco2}, \ref{fig:SHAPworking_capacity} and \ref{fig:SHAPh2o}) and a larger set of 100 descriptors composed by the previous and some additional (non-meaningful) ones.
%
%
In particular, we have chosen the non-relevant features among the ones discarded by AutoMatminer during its automatic feature reduction procedure, before the model training.
SL was adopted to find the maximum property value among a random subset of 500 samples from the original datasets (over 8000 MOFs), starting from a pool of 100 points with the lowest target property. 
Unexpectedly, SL optimization in the space of relevant descriptors does not ensure, in general, a faster convergence of the procedure to the optimum property value (this is also confirmed by results in the synthetic case reported in Supplementary Section S1).
Furthermore, among the three regression methodologies examined, only COMBO-based methods were able to provide always a faster convergence to the optimum value as compared to the random choice strategy.
Results are shown in Fig.~\ref{fig:SL}.

\subsection*{Optimization under incomplete access to the isosteric field of candidate MOFs-water working pairs}
\begin{figure}
\centering
\includegraphics[width=1.0\textwidth]{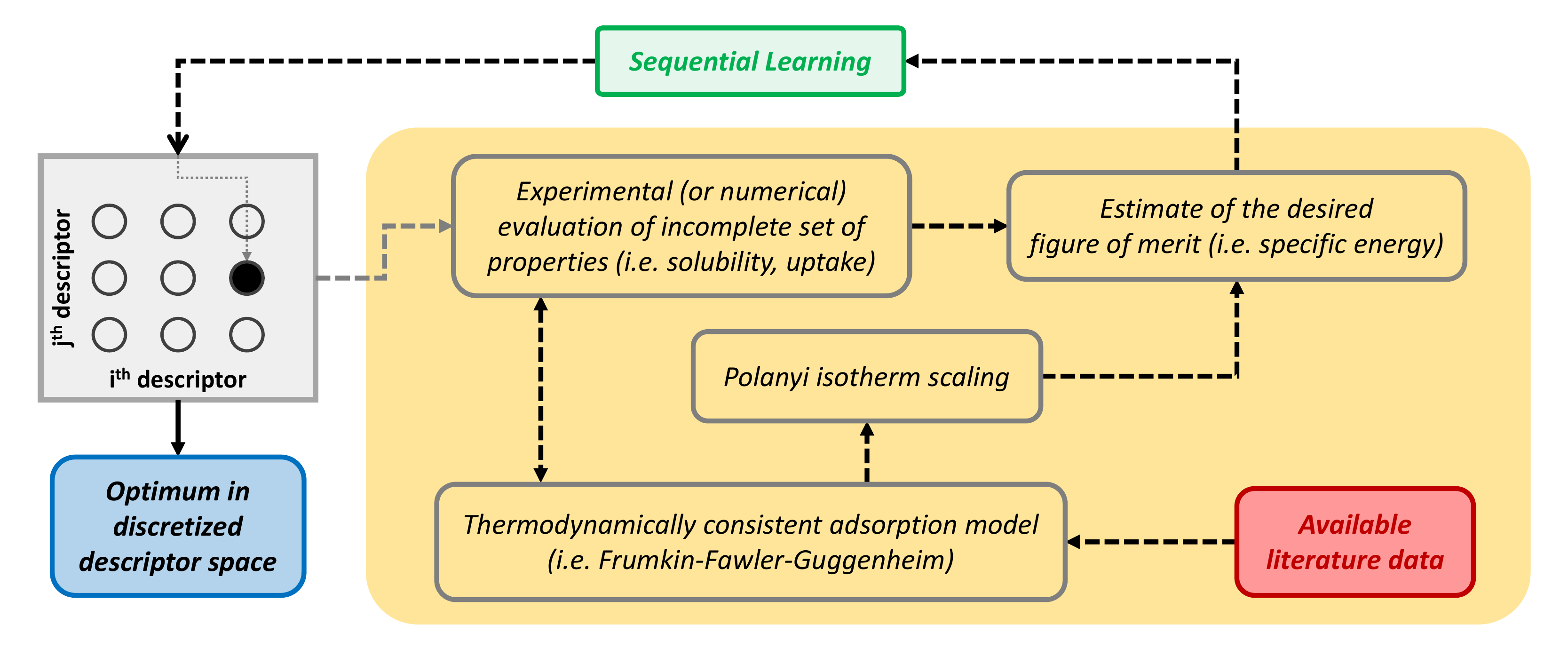}
\caption{Suggested procedure for estimating the specific energy of hypothetical MOF-water working pairs when only an incomplete knowledge of the isosteric field is experimentally or numerically accessible (highlighted in yellow).}
\label{fig:Schema}
\end{figure}
Without losing generality, here we aim at investigating the expected performance of hypothetical (i.e. computationally generated) MOFs for an important energy engineering application, namely water-sorption seasonal thermal energy storage (see also the Methods section).
The most challenging aspect of this task consists in the access to the entire isosteric field of each candidate MOF-water pair for estimating the engineering figure of merit of interest.
Clearly, for a large number of MOFs candidates, this is challenging and time consuming both computationally (typically, only the Henry low-coverage regime is reported in literature works \cite{wu2019understanding,yu2021efficient}) and experimentally \cite{lavagna2020cementitious}.
In this section, we specifically focus on such challenging aspect.
We envision an efficient optimization procedure that is capable of searching MOFs with the largest expected figure of merit of engineering relevance (either by SL, if materials are sequentially synthesized/computed, or by accessing readily available databases \cite{talirz2020materials}).
As far as seasonal thermal energy storage applications are concerned, here we focus on the highest specific energy of MOF-water working pairs among the compounds reported in ref. \cite{boyd2019data}. An overview of the proposed methodology is schematically reported in Fig. \ref{fig:Schema}.

%

As detailed in the Methods section, the ideal thermodynamic cycle of a closed sorption thermal energy storage system is completely defined by four operating temperatures.
In this study, we assume: $T_A=308\, \si{\kelvin}$ (the minimum temperature on the user side), $T_C=353\, \si{\kelvin}$ (the maximum temperature on the source side), $T_E=278\, \si{\kelvin}$ (the average winter temperature), $T_F=303\, \si{\kelvin}$ (the average summer temperature).
Those temperature values are reasonable for space heating applications in temperate climates \cite{lavagna2020cementitious}.
Given the Antoine equation, the equilibrium water vapour pressures $p_E = 866.2\, \si{\pascal}$ at the evaporator and $p_F=4231.6\, \si{\pascal}$ at the condenser are also uniquely defined considering the average winter and summer temperatures, respectively.
We decided to evaluate and maximize over the database one the most important engineering quantities in a thermal energy storage plant, namely the cycled heat per unit of material weight.
While full details on the adopted models are given in the Methods section, in the following we report and discuss the main simplifying assumptions in our approach:
%
%
\begin{itemize}
\item A key quantity to be estimated is the \ch{H2O}  working capacity. 
That quantity is related to the available adsorption sites $n_{TOT}$ per unit of dry sorbent mass.
Boyd \emph{et al.} \cite{boyd2019data} reported only the \ch{CO2} working capacity, while no data are available on the maximum \ch{H2O} uptake.
Nonetheless, we can rely on other related properties, such as the specific surface area of MOD.
To this end, we notice that Chaemchuen \emph{et al.} have reported \ch{H2O} working capacity for a pool of 66 MOFs \cite{chaemchuen2018tunable}.
A good correlation between the water uptake and the surface area (i.e., the available internal surface per gram of dry adsorbent) can be observed for typical MOFs used in the energy engineering field.
On the basis of that correlation, we impose a linear regression for finding the constant of proportionality between water uptake and surface area ($\textrm{water uptake}=\eta\times\textrm{surface area}$).
This yields $\eta=3.875\times10^{-4}\, \si{g_{\ch{H2O}}\per\meter\squared}$. 
More details can be found in Supplementary Section S4.
%

%
%
\item Henry coefficients $\widetilde{H}(T_0)$ for \ch{H2O} are listed at the reference temperature $T_0=298\, \si{\kelvin}$ with units of \si{\mol_{\ch{H2O}}\per\kilogram_{MOF}\per\bar}, thus representing the moles of adsorbed \ch{H2O} per kilogram of dry MOF per bar of  \ch{H2O} vapour. 
In our approach, we adopt the Frumkin-Fawler-Guggenheim (FFG) model to estimate the adsorption isotherm over the entire range of coverages only relying upon such Henry coefficient.
However, as discussed in the Methods sections, the FFG equation requires $H(T_0)$ in units of \si{\per\pascal}: we have thus converted $\widetilde{H}(T_0)$ (readily available from ref. \cite{boyd2019data}) to $H(T_0)$. 
Let $n_s$, $m_{MOF}$ and $p_{\ch{H2O}}$ be the number of adsorbed water moles, the mass of the potential MOF and the pressure of water in vapour phase, respectively, it holds:
\begin{equation}
\widetilde{H}(T_0)=\frac{n_s}{m_{MOF}p_{\ch{H2O}}}.
\end{equation}
The approximation of low-coverage regime yields the linear relationship between the coverage and pressure, namely  $\theta=H(T_0)p_{\ch{H2O}}$. Since the number of adsorbed water moles is related to the molar based total number of adsorption sites as $n_s=\theta n_{TOT}$, it follows:
\begin{equation}
H(T_0)=\widetilde{H}(T_0)\frac{\mathcal{M}_{MOF}}{n_{TOT}/n_{MOF}} ,
\label{eq:henry3}
\end{equation}
with $m_{MOF}=\mathcal{M}_{MOF}n_{MOF}$, $\mathcal{M}_{MOF}$ the molecular weight of the MOF, and $n_{MOF}$ the total number of moles.
Furthermore, the molar based total number of adsorption sites $n_{TOT}$ corresponds to the maximum number of water moles $n_{MAX,\ch{H2O}}$ that can be adsorbed, and the following relationship holds:
\begin{equation}
\frac{n_{MAX,\ch{H2O}}}{n_{MOF}}=\frac{m_{MAX,\ch{H2O}}}{m_{MOF}}\frac{\mathcal{M}_{MOF}}{\mathcal{M}_{\ch{H2O}}} ,
\label{eq:moles}
\end{equation}
where $\mathcal{M}_{\ch{H2O}}$ is the molecular weight of water and $m_{MAX,\ch{H2O}}$ denotes the maximum mass of water that can be adsorbed. 
The ratio $m_{MAX,\ch{H2O}}/m_{MOF}$ is related to the \ch{H2O} working capacity of the MOF and it is equal to $\eta S$, where $\eta$ is the constant of proportionality between the uptake and the surface area $S$.
A comparison of Eq.~\ref{eq:moles} and Eq.~\ref{eq:henry3} yields:
\begin{equation}
H(T_0)=\widetilde{H}(T_0)\frac{\mathcal{M}_{\ch{H2O}}}{\eta S}\times10^{-8} ,
\label{eq:conversion}
\end{equation}
which is the final conversion formula of the Henry coefficient for \ch{H2O} from measure units of \si{\mol_{\ch{H2O}}~\kilogram_{MOF}^{-1}~\bar^{-1}} into \si{\per\pascal}. 
Here, the factor $10^{-8}$ appears because $[\mathcal{M}_{\ch{H2O}}]=\, \si{\gram_{\ch{H2O}}\per\mol_{H2O}}$, $[\eta]=\, \si{\gram_{\ch{H2O}}\per\meter\squared}$, $[S]=\, \si{\meter\squared\per\gram_{MOF}}$, and so $[\widetilde{H}(T_0)\mathcal{M}_{\ch{H2O}}/(\eta S)]=\, \si{\gram_{MOF}~\kilogram_{MOF}^{-1}~\bar^{-1}}$.

\item A crucial quantity for heat transformation is the isosteric heat of adsorption $q_{st}$. 
Due to the Clausius-Clapeyron relationship (see eq. \ref{ClausiusClapeyron} in the Methods section), at least two adsorption isotherm curves (at $T_A$ and at $T_C$) are needed to estimate the corresponding $q_{st}$.
In our database, though, the Henry coefficients are only available at $T_0$.
In order to reconstruct a second adsorption isotherm for the same MOF-water working pair, we decided to resort to the potential theory of Polanyi, thus exploiting the basic notion that all adsorption isotherms are self-similar when rescaled with respect to the Polanyi potential function. 
Details are provided in the Supplementary Section S5.
More specifically, the Polanyi potential is defined as:
\begin{equation}
\mathcal{A} = -RT\ln\left(\frac{p_s(T)}{p}\right),
\end{equation}
where $p_s(T)$ is the saturation pressure of water at temperature $T$, while $p$ is the pressure of the vapour phase on the adsorbent surface \cite{butt2013physics}.
Since, at a given pressure $p$, the Polanyi potential is a constant of the sorption pair, we have computed $\mathcal{A}$ at $T_0$ in a range from $10^{-4}\, \si{\pascal}$ up to $p_s(T_0)=3157\, \si{\pascal}$ (Antoine equation for water); then, in the $\theta - p$ chart, we have rescaled the abscissa $p$ of the isotherm obtained at the temperature $T_0$ according to $p=p_s(T)\exp\left(\mathcal{A}/(RT)\right)$, for getting the new curves at $T_A$ and $T_C$.
We have finally computed the isosteric heat by means of the Clausius-Clapeyron relationship:
\begin{equation}
q_{st}=\frac{R}{3}\frac{T_CT_A}{T_C-T_A}\sum_{i=1}^{3}\ln {\frac{p_2(\theta_i)}{p_1(\theta_i)}} ,
\end{equation}
where the points 1 and 2 represent the intersections of an isosteric transformation with the two isotherms respectively at $T_A$ and $T_C$. 
We have repeated the procedure for three coverage values (i.e. $\theta_1=0.4$, $\theta_2=0.5$, $\theta_3=0.6$) and averaged them.

\item Finally, upon determination of the low- and high-temperature adsorption isotherms curves at $T_A$ and $T_C$, the coverage span $\Delta\theta$ during the discharge phase can be determined as detailed in the Methods section. 
As for the estimate of the isosteric heat, this requires for all the compounds the rescaling of the horizontal axis in the $\theta - p$ chart of the isotherm obtained at temperature $T_0$ according to the Polanyi potential theory.
\end{itemize}

\begin{figure}
\centering
\includegraphics[width = 0.7\textwidth]{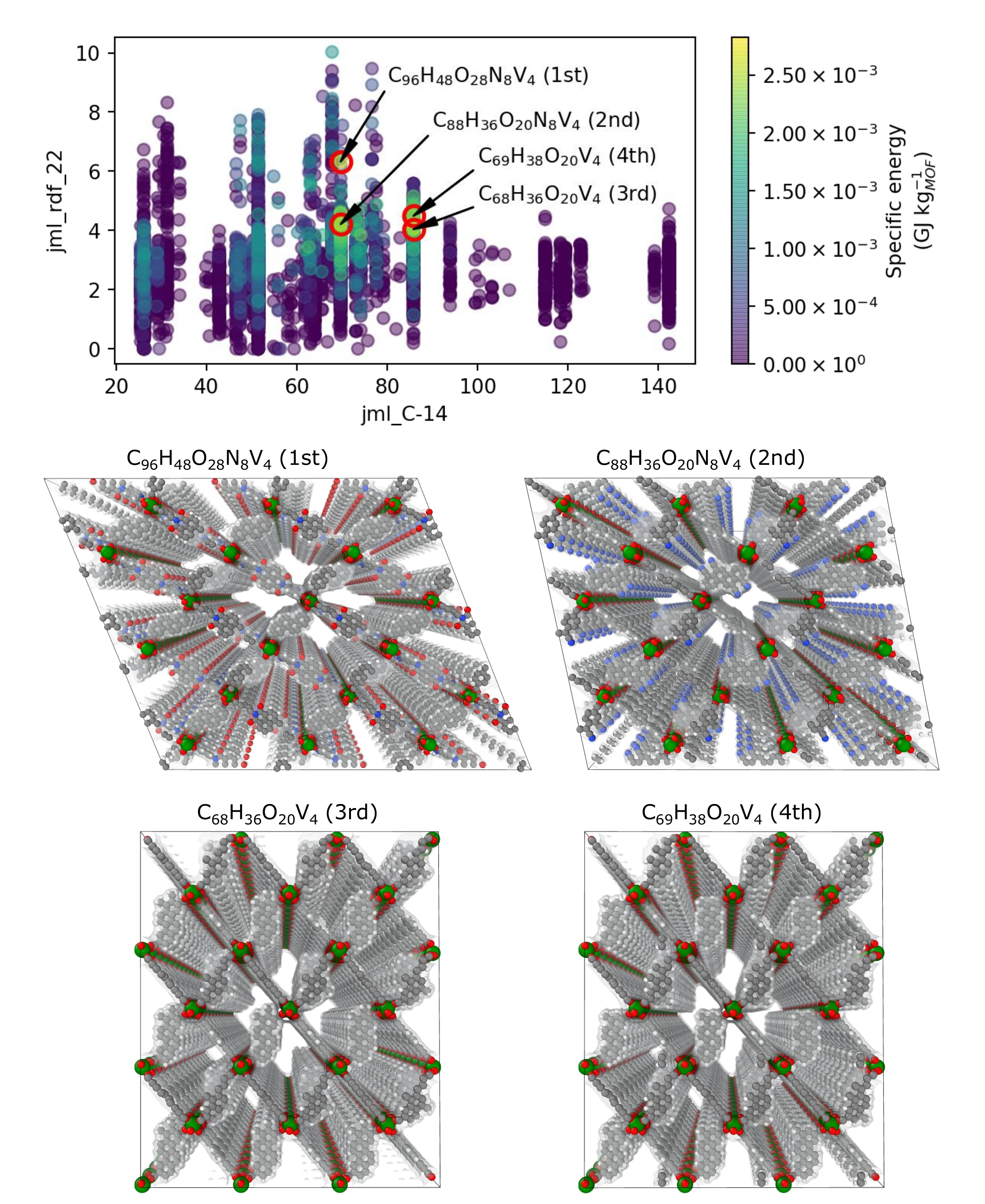}
\caption{The entire set of 5028 potential MOFs in the database by Boyd \emph{et al.} \cite{boyd2019data} is displayed in the 2D chart, where the first two SHAP ranked descriptors for the \ch{H2O} Henry coefficient are represented. The best four MOFs sorbents for the adsorption/desorption based thermal storage application are highlighted in the 2D chart, and 3x3x13 replications of the respective crystallographic cells depicted (C atoms: gray; H atoms: white; O atoms: red; N atoms: blue; V atoms: green).}
\label{fig:bestMOF}
\end{figure}

\begin{figure}
\centering
\includegraphics[width=0.5\textwidth]{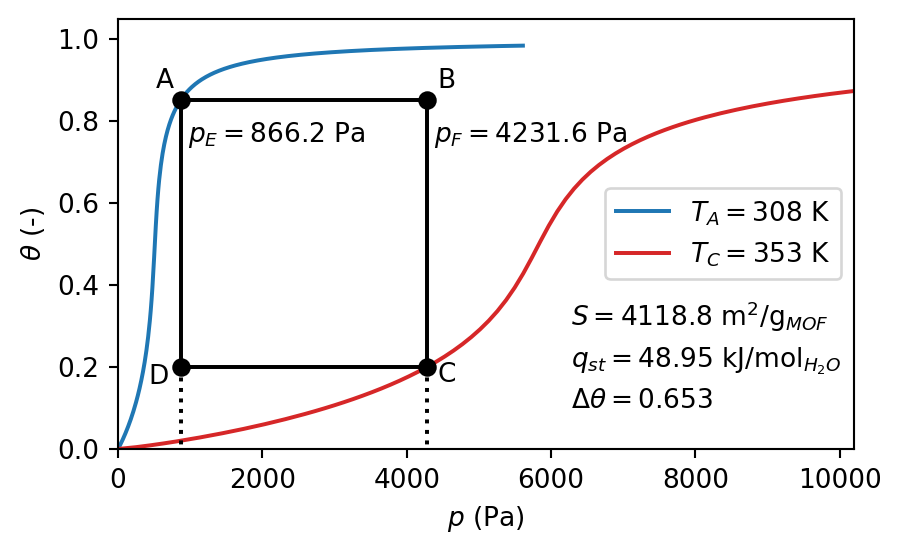}
\caption{Adsorption/desorption based thermal energy storage cycle for the potential MOF ``str\_m5\_o18\_o28\_sra\_sym.72'' with water, in the coverage-pressure plane. The isotherms $T_A=308\, \si{\kelvin}$ and $T_C=353\, \si{\kelvin}$ are shown. Surface $S$, isosteric heat $q_{st}$, and coverage span $\Delta\theta$ over the cycle are reported, giving an objective function $Sq_{st}\Delta\theta=1.32\times10^5\, \si{\kilo\joule~\meter\squared~\mol_{\ch{H2O}}^{-1}~\gram_{MOF}^{-1}}$ or equivalently $2.83\times10^{-3}\, \si{G\joule\per\kilogram_{MOF}}$, which corresponds to $1.73\, \si{GJ/m^3}$.}
\label{fig:optimum}
\end{figure}

We have thus computed the following objective function $Sq_{st}\Delta\theta$, which recovers the cycled heat up to a constant, over the entire list of 5028 potential MOFs with positive surface area in ref. \cite{boyd2019data}. 
The best MOF turned out to be the compound with chemical formula \ch{C96H48O28N8V4} and referred to as "str\_m5\_o18\_o28\_sra\_sym.72" in the database (see also the molecular rendering in Fig.~\ref{fig:bestMOF}), with Henry coefficient at $298\, \si{\kelvin}$ of $6110.54\, \si{\mol_{\ch{H2O}}~\kilogram_{MOF}^{-1}~\bar^{-1}}$ (or equivalently, $6.89\times10^{-4}\, \si{\per\pascal}$) and surface area $S=4118.79\, \si{\meter\squared\per\gram_{MOF}}$. 
Figure \ref{fig:optimum} shows also the ideal expected thermodynamic cycle related to this optimal potential MOF.

We observe a coverage span $\Delta\theta=0.653$, an isosteric heat $q_{st}=48.95\, \si{\kilo\joule\per\mol_{\ch{H2O}}}$ with an objective function value of $1.32\times10^{5}\, \si{\kilo\joule~\meter\squared~\mol_{\ch{H2O}}^{-1}~\gram_{MOF}^{-1}}$.
That quantity can be directly related to specific energy: upon multiplication by the constant $\eta/\mathcal{M}_{\ch{H2O}}$ ($\eta = 3.875\times10^{-4}\, \si{g_{H_2O}/\meter\squared}$, $\mathcal{M}_{\ch{H2O}} = 18.02\, \si{\gram_{\ch{H2O}}/\mol_{\ch{H2O}}}$), we obtain a value of $2.83\times10^{-3}\, \si{G\joule\per\kilogram_{MOF}}$.
Furthermore, we can compute the theoretical density $\rho_{MOF}$ of the crystal knowing the mass of the cell ($1965.21\, \si{u}=3.263\times10^{-21}\, \si{\gram_{MOF}}$, as from the database) and its volume ($5.335\times 10^{-21}\, \si{cm^{3}}$, as from the CIF file), leading to $\rho_{MOF}=0.612\, \si{\gram_{MOF}\per\cm^{3}}$. 
As a result, the volume-based energy density turns out to be $1.73\, \si{GJ/m^3}$.
For the sake of comparison, Fig.~\ref{fig:bestMOF} shows a 2D map where the two axes represent the two most important descriptors according to the SHAP ranking for the Henry coefficient of \ch{H2O}: the four top performing potential MOFs are highlighted and the corresponding crystallographic cells depicted. Moreover, Table~\ref{tab:MOFs} shows the ten most performing potential MOFs ranked in terms of specific energy.

\begin{figure}
\centering
\includegraphics[width=.90\textwidth]{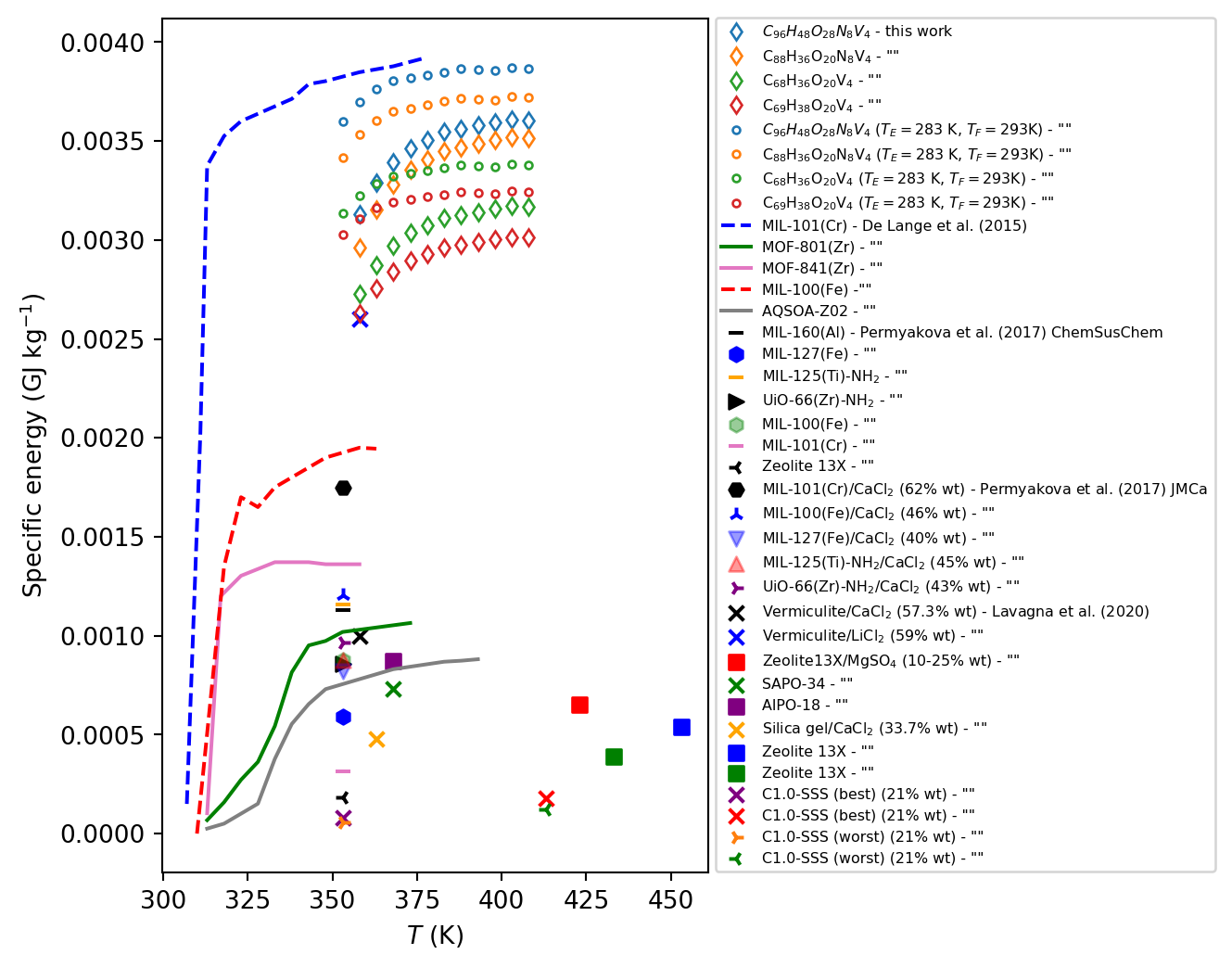}
\caption{Comparison between the expected specific energy for different desorption temperatures of the optimum MOFs identified in this work (either standard environmental conditions, i.e. evaporation temperature $T_E = 278\, \si{\kelvin}$ and condensation temperature $T_F = 303\, \si{\kelvin}$, or with conditions of $T_E = 283\, \si{\kelvin}$ and $T_F = 293\, \si{\kelvin}$) and several water sorbent materials reported in the literature. 
}
\label{fig:Comparison}
\end{figure}

As depicted in Fig. \ref{fig:Comparison}, the four top performing potential MOFs show (material based) specific energy values among the highest available in the literature for sorption thermal energy storage under similar operating conditions.

\begin{table}[h]
\centering
\caption{\label{tab:MOFs} Top ten potential MOFs in terms of specific stored energy from the database by Boyd \emph{et al.} \cite{boyd2019data}. The name indicated in the database, the brute formula, the molecular weight, the surface area and the specific energy are shown.}
\begin{tabular}{llccc}
\toprule
Database name & Brute formula & Molecular weight & Surface area & Specific energy\\
&& (u) &($\si{m^2/g_{MOF}}$) & ($\times 10^{-3}\, \si{\giga\joule\per\kilogram_{MOF}}$)\\
\midrule
``str\_m5\_o18\_o28\_sra\_sym.72"& \ch{C96H48O28N8V4} & 1965.21 & 4118.79 & 2.83\\
``str\_m5\_o3\_o18\_sra\_sym.73" & \ch{C88H36O20N8V4} & 1729.03 & 3888.89 & 2.57\\
``str\_m5\_o6\_o18\_sra\_sym.82" & \ch{C68H36O20V4} & 1376.76 & 3577.31 & 2.44\\
``str\_m5\_o6\_o18\_sra\_sym.92" & \ch{C69H38O20V4} & 1390.79 & 3473.55 & 2.40\\
``str\_m5\_o7\_o18\_sra\_sym.115" & \ch{C76H32O20N4V4} & 1524.85 & 3428.98 & 2.36 \\
``str\_m5\_o7\_o18\_sra\_sym.133" & \ch{C72H36O28V4} & 1552.81 & 3384.12 & 2.34 \\
``str\_m5\_o7\_o18\_sra\_sym.136" & \ch{C72H40O20N4V4} & 1484.87 & 3362.19 & 2.31 \\
``str\_m5\_o7\_o18\_sra\_sym.20" & \ch{C76H36O24V4} & 1536.85 & 3340.13 & 2.22 \\
``str\_m5\_o7\_o18\_sra\_sym.124" & \ch{C72H40O20N4V4} & 1484.87 & 3355.61 & 2.22 \\
``str\_m5\_o6\_o18\_sra\_sym.17" & \ch{C70H40O19V4} & 1404.82 & 3189.28 & 2.22 \\

\bottomrule
\end{tabular}
\end{table}

\clearpage

\section*{Discussion}
Sequential Learning (SL) algorithms can in principle dramatically reduce the number of evaluations needed for finding the optimum of an unknown function as compared with a naive random choice and, as such, they are emerging as effective tools for material optimization and discovery.
In this work, focusing on Metal Organic Frameworks - MOFs and some of their crucial adsorption properties (both with \ch{H2O} and \ch{CO2} as sorbate fluids), we have addressed a number of critical aspects related to the discovery of the minimal set of important crystallographic descriptors for SL based optimization algorithms.
We have shown that the general protocol for sorting out the minimal set of ruling descriptors (here referred to as crystallographic {\it genetic code}) for a given adsorption property is based on two steps: i) construction and training of Machine Learning (ML) model which identifies the number of ruling descriptors; ii) evaluation of the relative importance of each explanatory variable on the chosen output by the SHAP analysis.
%
We found that, as long as the set of such ruling descriptors (for a given property of interest) is included among the exploration space features, convergence performance is not affected, although the computational burden of a SL algorithm also depends on the dimension of the parameter space to be explored: taking into account only the most relevant features may be in fact beneficial in that respect.
Furthermore, based on the several examples provided here (i.e. Henry coefficient for \ch{CO2}, Henry coefficient for \ch{H2O}, working capacity for \ch{CO2} as well as the synthetic example discussed in the Supplementary Section S1), we have consistently noticed that only the COMBO algorithms always perform better than random guessing.

Furthermore, we recognize that full access to the adsorption properties of hypothetical MOFs in the entire coverage regime (as requested in important applications of engineering relevance) is very challenging both experimentally and computationally.
This holds particularly for water-MOFs working pairs that are promising for a number of energy applications.
Hence, we formulate an innovative, general and efficient computational screening of hypothetical MOFs which, only relying upon the adsorption properties reported in Fig. \ref{fig:models}, is capable to estimate important figures of merit for sorption based seasonal thermal energy storage. 
Remarkably, our procedure suggests that some of the MOFs hypothesized in the database by ref. \cite{boyd2019data} (developed for completely different purposes) are expected to largely outperform most of the state-of-the-art water sorbent compounds.
We believe that the above results represent a first important step towards efficient MOFs screening and optimization, not only with respect to intrinsic materials properties but also (and importantly) with respect to figures of merit of engineering relevance for applications such as thermally driven water harvesting from air, water sorption thermal energy storage, and solar cooling.

Clearly, our approach is based on a number of approximations and it still requires additional research activities, that are currently ongoing.
First, we notice that a large set of hypothetical MOFs may be characterized by properties (e.g. Henry coefficients) whose values span several orders of magnitude.
Hence a unique ML model, as used in this pipeline, may achieve a high coefficient of determination if its logarithm is considered.
Nonetheless, the computation of the coverage span $\Delta\theta$ depends directly on the Henry coefficient, which may thus be affected by a relatively high error.
Furthermore, additional simplifying assumptions that have been used in our approach include fixed parameters such as the constant of proportionality between the specific surface area and the water working capacity, as well as the {\it steepness} coefficient $\beta$ in the FFG model.
Without loosing generality, those assumptions could be relaxed in the near future relying on more sophisticated models.
One possible way to tackle those challenges (not necessarily the only possible strategy) may be a preliminary classification of the hypothetical MOFs based on properly trained ML classifiers, with the purpose of assigning a given compound of interest to a specific category (e.g. the set of MOFs with Henry coefficient of similar magnitude, similar $\beta$, etc.), then providing property predictions on each MOF category.
%

\section*{Methods\label{sec:level2}}

\subsection*{Water sorption thermal energy storage\label{sec:TES}}

In this work we focus on the use of potential MOFs for water-sorption based thermal storage applications. 
%
%
Physical adsorption processes are based on weak and reversible interactions between the (solid) sorbent material and the corresponding adsorbate, i.e. the fluid \cite{ruthven1984principles}.
Those phenomena are relevant to thermal energy engineering as sorption/desorption in solid sorbents can be accompanied by significant amount of energy exchange. In the following, the solid sorbent are MOFs, while water is the adsorbate.

To allow desorption of an infinitesimal number ($\textrm{d}n$) of moles of adsorbate from the adsorbent surface, a given amount of heat $\textrm{d}Q = q_{st}\,\textrm{d}n$ has to be provided to the system, where $q_{st}$ (with units of \si{k\joule\per mol}) denotes the isosteric heat. 
Since the process is reversible, the same amount of heat $\textrm{d}Q$ is released by the dry sorbent when $\textrm{d}n$ moles of fluid at a pressure $p$, initially in vapour phase, are adsorbed.
Furthermore, we define the load $X$ as the ratio between the mass of adsorbate and the mass of dry sorbent. 
A process characterized by constant load $X$ is referred to as an \emph{isosteric transformation} and the popular Clausius-Clapeyron relationship yields:
\begin{equation}\label{ClausiusClapeyron}
\left(\frac{\partial\ln{p}}{\partial\left(-\frac{1}{T}\right)}\right)_X = \frac{q_{st}}{R},
\end{equation}
where $T$ is the absolute temperature and $R=8.314\, \si{\joule\per\mol\per\kelvin}$ denotes the gas constant \cite{schmidt2004optimizing}.
Therefore, an isosteric transformation in the Clapeyron chart ($\ln{p}$ vs $-1/T$) is a curve with local slope $q_{st}/R$. 
Similarly, the adsorbate isosteric curve has a slope $\Delta H_{(vap)}/R$, where $\Delta H_{(vap)}$ is the molar enthalpy for liquid-vapour phase change of the adsorbate.

Closely related to the load $X$, the coverage $\theta$ is defined as the ratio between the number of already occupied adsorption sites $n_s$ and the total available number of sites $n_{TOT}$. 
At equilibrium and at a given temperature $T$, the coverage $\theta$ depends on the pressure $p$ of the vapour phase according to an adsorption isotherm, whose shape depends on the sorbent/adsorbate pair. 
MOFs/water pairs are known to show typical ``S-shaped" isotherms in the $\theta-p$ chart (i.e. type V of the IUPAC classification \cite{sing1985reporting}).
Thus, in the following, we make the assumption that the Frumkin-Fawler-Guggenheim (FFG) model can be used conveniently for describing analytically the MOFs-water adsorption isotherms: %
\begin{equation}
\theta = \frac{H(T)p\exp{(\beta\theta)}}{1+H(T)p\exp{(\beta\theta)}} ,
\label{eq:theta}
\end{equation}
where $\beta=\frac{\overline{n}E_p}{RT}$ rules the \emph{steepness} of the ``S-shape'', $\overline{n}$ denotes the neighboring binding sites and $E_p$ represents the additional binding energy due to lateral interactions \cite{butt2013physics}. 
We have used the FFG model to interpret eight experimental isotherms of real MOF-water pairs and achieve a proper choice of $\beta$. In particular, for each curve, we have identified the best value of $\beta$ in terms of a least squares approach; then, we have taken the mean over those 8 values, ending up with $\beta = 3.4$. More details can be found in the Supplementary Section S6. Finally, $H(T)$ is the Henry coefficient for the specific sorbent/adsorbate pair (with units \si{\per\pascal}) at a given absolute temperature $T$.

%

\begin{figure*}[ht]
\centering
\includegraphics[width=0.8\textwidth]{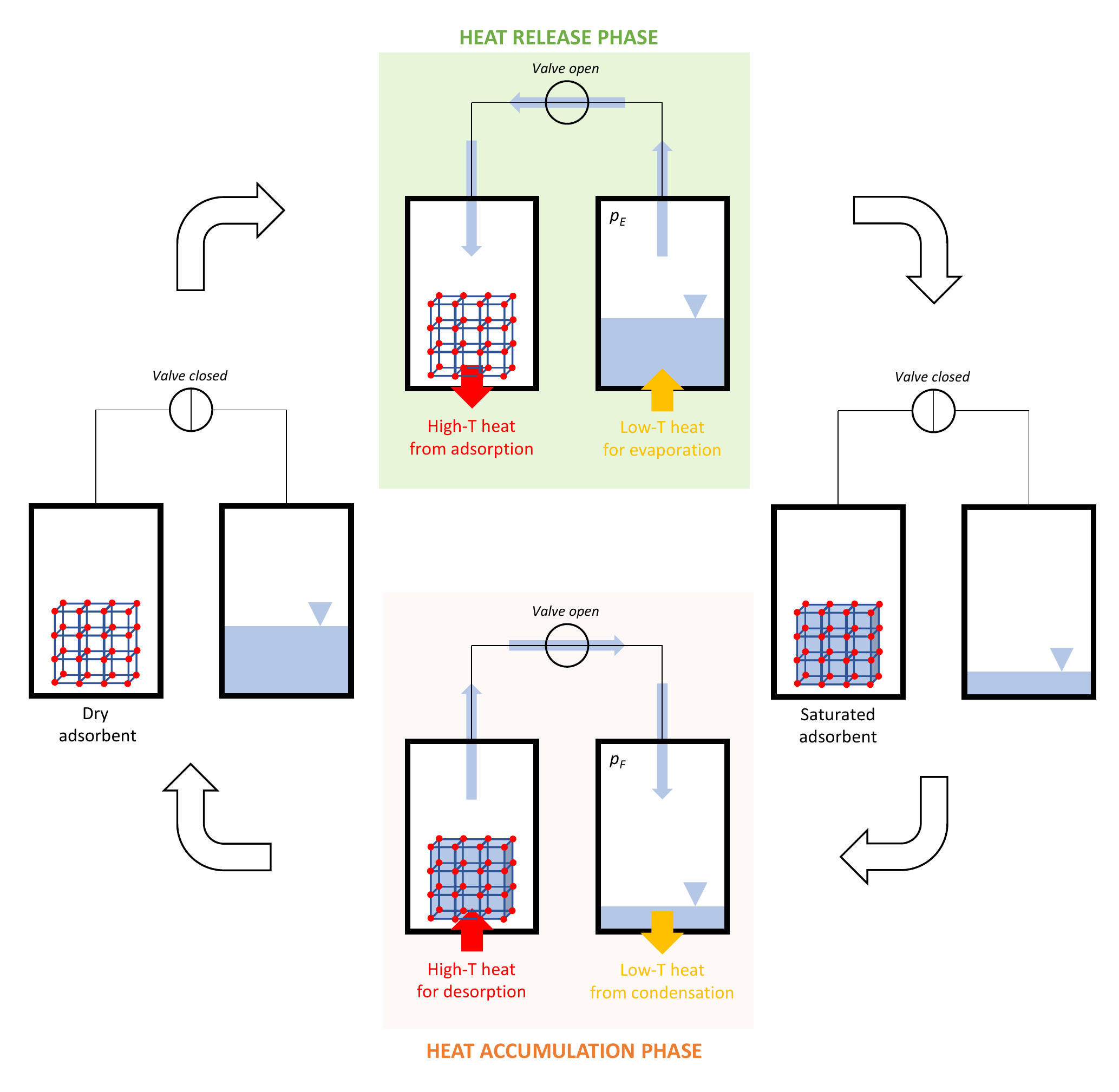}
\caption{Schematic of a closed water sorption thermal energy storage system, in which the cycle underlying the heat accumulation and successive release is represented.}
\label{fig:HP_scheme}
\end{figure*}

\begin{figure}
    \centering
    \includegraphics[width=0.5\textwidth]{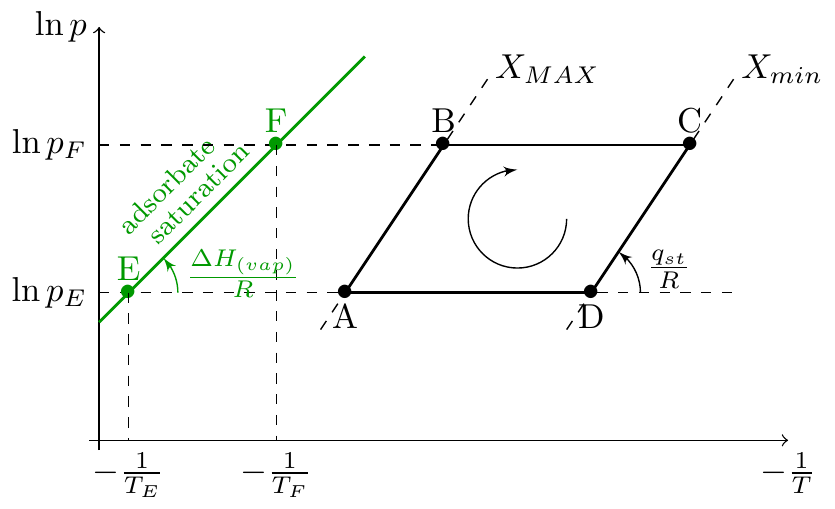}
    \includegraphics[width=0.5\textwidth]{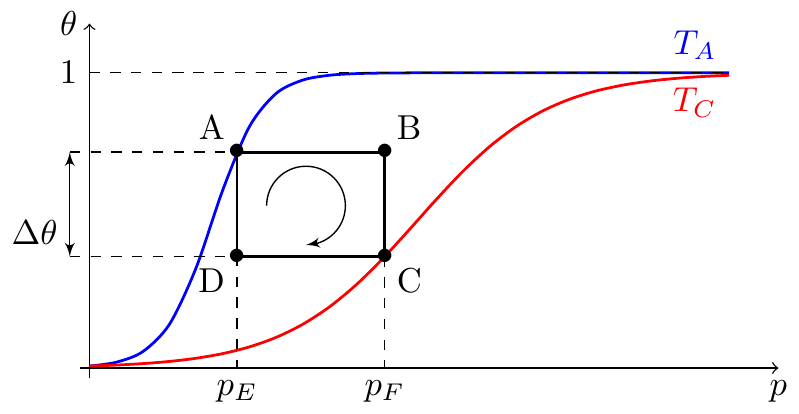}
    \caption{Schematics of ideal adsorption/desorption thermal energy storage cycle. Top: Ideal cycle in the Clapeyron chart. Bottom: Same ideal cycle in the coverage-pressure chart, between the two limiting isotherms passing by $\theta(p, T_A)$ and $\theta(p, T_C)$.}
    \label{fig:cycle}
\end{figure}

%
A schematic of a closed water sorption thermal energy storage system is shown in Fig.~\ref{fig:HP_scheme}. These systems are based on a reactor, containing the solid sorbent, connected with a condenser/evaporator by means of a valve \cite{fasano2016}. This chemical apparatus follows a seasonal closed cycle completely defined by four temperatures: $T_A$ (the minimum temperature on the user side), $T_C$ (the maximum temperature on the source side), $T_E$ (the average winter temperature), $T_F$ (the average summer temperature). 
The two pressures $p_E$ (evaporator) and $p_F$ (condenser) are related to the absolute temperatures $T_E$ and $T_F$, respectively, through the Antoine equation for water saturation $p_{E,F} = 133.2\times10^{A-B/(C+T_{E,F}-273)}$, where $A=8.07131$, $B=1730.63$, $C=233.426$ \cite{dean1990lange}.
Hence, the ideal thermodynamic cycle of a closed thermal energy storage process (see Fig.~\ref{fig:cycle}) is based on the following four steps:
\begin{enumerate}
\item The sorbent/adsorbate is heated isosterically up to a temperature $T_B$, corresponding to a pressure $p_F$ in the condenser (line AB).
\item Heating of the pair continues at constant pressure $p_F$ and desorbed vapour flows to the condenser through the opened valve. 
In the condenser, the adsorbate rejects the condensation heat into the environment while condensing until the maximum temperature of the heat source $T_C$ is reached (line BC). 
The condition of minimum load is reached and the valve gets closed.
\item Keeping the valve closed, the system in contact with the environmental temperature cools isosterically during the storage period (line CD) down to temperature $T_D$, corresponding to the evaporator pressure $p_E$.
\item During the discharge phase of the heat storage system, the valve is opened to let the adsorbate evaporate and reach the reactor. 
During this isobaric transformation (line DA), the heat of adsorption $Q_{DA}$, also known as cycled heat, is released. 
\end{enumerate}

One of the most important figures of merit for energy storage systems is the volumetric energy density, namely the maximum energy that can be stored per unit of volume of the plant \cite{dincer2021thermal}.
Clearly, at fixed plant size, the higher the cycled heat the higher the volumetric energy density of the storage system. In this view, the choice of the solid sorbent material for a given adsorbate is key for maximizing the cycled heat in the ideal thermodynamic cycle and therefore the expected volumetric energy density. 
We thus perform below a material screening aiming at the maximum value of the following cycled heat (i.e., the released heat during the DA process in Fig. \ref{fig:cycle}):
\begin{equation}\label{cycledheat}
Q_{DA}=\int_D^Aq_{st}\,\textrm{d}n_s = \int_D^An_{TOT}q_{st}\,\textrm{d}\theta \approx n_{TOT}q_{st}\Delta\theta ,
\end{equation}
where we have used the definition of heat of adsorption $\textrm{d}Q = q_{st}\,\textrm{d}n_s$, coverage $\theta = n_s/n_{TOT}$ and the approximation $q_{st}\approx\textrm{const}$.

\subsection*{Sequential Learning}
%

Typical steps in any SL algorithm consist in (i) constructing a regression model over known data, (ii) using a strategy to suggest the best-unmeasured point to test, (iii) enlarging the known dataset with this tested point and (iv) iterating up to the tested candidate meets the needed specification.
Let $\mathcal{D}=\{(\mathbf{x}_1,y_1),\dots, (\mathbf{x}_n,y_n)\}$ denotes a set of $n$ training data, where $\mathbf{x}_i\in\mathbb{R}^d$ and $y_i\in\mathbb{R}$ represent the $i$-th vector of descriptors and its known response, respectively.
Let $\{\mathbf{x}_{n+1},\dots,\mathbf{x}_m\}$ denote the $m-n$ $d$-dimensional arrays of descriptors with unknown responses $\{y_{n+1},\dots,y_m$\}. 
To find the location $\mathbf{x}^{*}$ of the maximum $y^{*}$, we would need the exact model $y=f(\mathbf{x})$.
However, given the restricted set $\mathcal{D}$ of training data, only a surrogate model $y=\hat{f}(\mathbf{x}|\mathcal{D})$ can be constructed. 
Hence, for each unmeasured point $i=n+1,\dots,m$, different regression methodologies (here FUELS-Random Forest, kriging and COMBO-Gaussian processes) can be used to estimate the response $\hat{f}(\mathbf{x}_i)$ in terms of a mean value $\mu(\mathbf{x}_i)$ and the corresponding uncertainty $\sigma(\mathbf{x}_i)$, indicating the robustness of the prediction. 
To measure the performance of any combination \emph{regression model/query strategy}, we put ourselves in the practitioner's perspective, who is interested in a unique sequence of points to be tested, and not in an average over more paths (as, for instance, shown in ref. \cite{ling2017high}). To achieve this, for those regression models not allowing a deterministic prediction (i.e., Random Forest and COMBO), at each step we have repeated $100$ times the choice of the next point to query, picking the most preferred one. 
A comprehensive comparison of the above methodologies is reported in the results, and the complete details on the adopted algorithms can be found in the Supplementary Sections S7 and S8.

\subsection*{Model training and choice of the descriptors}
The first issue to be addressed when applying SL to material optimization is computation and selection of relevant features (or descriptors).
The descriptor issue is critical in materials science \cite{ghiringhelli2015big,gomes2021machine} as well as in other computational fields \cite{chiavazzo2017intrinsic}.
In this work, we first investigate to which extent the choice of a minimal set of relevant descriptors is critical for the fast convergence of SL algorithms.      

To this end, before even implementing SL procedures, we decided to perform a preliminary feature \emph{pruning} for discovering the most meaningful ones in terms of the target property.
We use the entire dataset (both descriptors and target property) to train and validate a regression model by means of AutoMatminer \cite{dunn2020benchmarking}, which allows to automatically train and validate a complete pipeline - feature reduction, data cleaning and machine learning - with automatic hyper-parameter tuning. In particular, AutoMatminer contains a customized dictionary of operators in terms of feature preprocessors, feature selectors and ML models. The algorithm automatically searches for their best combination, with the best ML model hyperparameters. Specifically, we have removed some of the default operators from the AutoMatminer proposed list, in order to have a stable code (e.g., to avoid the drop of all the features in the preprocessing step). The list of the retained ones can be found in the Supplementary Section S3. 
The degree of accuracy and the training time depend on the choice of the predefined set of options, which specify how the search of the best pipeline works.
We choose the preset ``express'' for a synthetic dataset (moderate accuracy and relatively quick training), and the preset ``production'' for the effective MOFs datasets (higher accuracy and slower training).
Upon model training and validation, we detect the most important features thanks to the Kernel SHAP algorithm in its model agnostic realization \cite{NIPS2017_7062}, thus quantifying to which extent a given feature impacts on the output.
The latter methodology is based on the classical Shapley value, which has in game theory its original field of application. There, the problem of assigning, in a cooperative game, a proportional reward to each player is addressed based on the real contribution provided to the common objective of the coalition. 
In a model, given $F$ the set of all features and its generic subset $S\subseteq F$, the importance of the $i-\textrm{th}$ descriptor depends on the comparison between the model $f_{S\cup \{i\}}$ trained with that explanatory variable, and another model $f_S$ trained without that feature; then, the difference between the predictions $f_{S\cup\{i\}}(\mathbf{x}_{S\cup\{ i\}})-f_S(\mathbf{x}_S)$ is computed, where $\mathbf{x}_{S\cup\{i\}}$ and $\mathbf{x}_S$ represent respectively the values of the input features over the subsets $S\cup\{i\}$ and $S$. This difference is weighted over all possible subsets $S$ and the importance value of the $i-\textrm{th}$ feature turns out to be
\begin{equation}
\phi_i = \sum_{S\subseteq F\setminus \{i\}}\frac{|S|!(|F|-|S|-1)!}{|F|!}\left(f_{S\cup\{i\}}(\mathbf{x}_{S\cup\{ i\}})-f_S(\mathbf{x}_S)\right),
\label{eq:SHAP}
\end{equation}
where $|\cdot|$ denotes the number of elements. Because of the huge number of possible descriptors subsets $S$ of a set $F$, the classical Shapley values of Eq.~\ref{eq:SHAP} are computationally challenging: therefore, the SHAP package provides suitable approximations.

\section*{Data availability}
The data that support the findings of this study are available from the corresponding author upon reasonable request.

\section*{Code availability}
Upon request to the corresponding Author.

\bibliographystyle{ieeetr}
\bibliography{biblio}

\section*{Acknowledgments}
E.C. acknowledges financial support of the Italian National Project PRIN \emph{Heat transfer and Thermal Energy Storage Enhancement by Foams and Nanoparticles} (2017F7KZWS) and of the research contract PTR 2019/21 ENEA ({\it Sviluppo di modelli per la caratterizzazione delle proprietà di scambio termico di PCM in presenza di additivi per il miglioramento dello scambio termico}) funded by the Italian Ministry of Economic Development (MiSE).

\section*{Authors contributions}
E.C. conceived the idea and found financial support. G.T. performed all computations and wrote the first paper draft. E.C., M.F and L.B. supervised the research activities. L.B. and M.F. helped with result presentation and interpretation. M.F. suggested the synthetic dataset. All authors contributed to final paper writing.

\section*{Competing interests}
The authors declare no competing interests.

\end{document}


\maketitle

\subsection*{S1: Synthetic dataset}

%
We have generated a synthetic dataset of 1000 samples and 20 columns by means of the ``Friedman \#1" regression problem \cite{friedman1991multivariate, breiman1996bagging, sklearnfriedman}, in which the target $y$ is given by the combination of five independent variables ($\mathbf{x}\in\mathbb{R}^5$):
\begin{equation}
y(\mathbf{x})=10\sin(\pi x_1 x_2) + 20\left(x_3-\frac{1}{2}\right)^2+10x_4+5x_5+\epsilon\mathcal{N}(0, 1) 
\end{equation} 
choosing a zero-centered and unit-variance noisy term normally distributed $\mathcal{N}(0,1)$, with $\epsilon=0.1$. More specifically, all the descriptors are independent and uniformly distributed over the interval $[0, 1]$, but only 5 of them are effectively used to compute the output. Therefore, the remaining 15 columns represent \emph{fake} features, with the only purpose of disturbing the regression model. 

\begin{figure}
\centering
\includegraphics[width=0.35\textwidth]{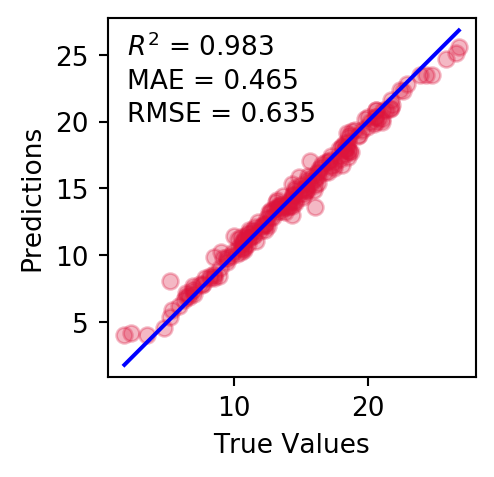}
\caption{Predictions of the regression model trained with AutoMatminer (preset ``express" of the ``Friedman\#1" problem) over the testing set. Model performance measures are shown in terms of coefficient of determination $R^2$, Mean Absolute Error (MAE), Root Mean Squared Error (RMSE).}
\label{friedman1}
\end{figure}
\begin{figure*}
\centering
\includegraphics[width=.75\textwidth]{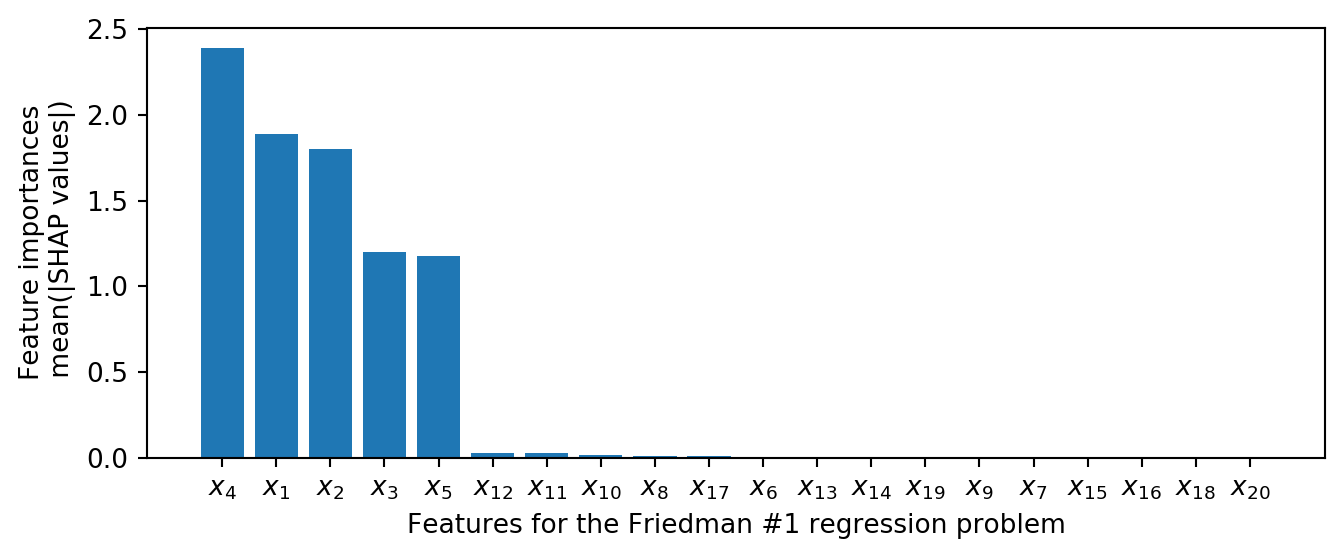}
\caption{Average impact of the 20 features on the model output for the ``Friedman\#1" regression problem over the whole testing set (200 samples) by Kernel SHAP, which recognizes correctly that only features $(x_1,\dots,x_5)$ are relevant.}
\label{friedman2}
\end{figure*}

For the identification of the important features, we have applied the model-agnostic interpretability algorithm of Kernel SHAP on a working regression model constructed with AutoMatminer. In particular, AutoMatminer trained a model on the 80\% of the 1000 samples by means of the ``express'' preset and tested it over the remaining 20\%, getting a regression performance of $R^2=0.978$ (Fig.~\ref{friedman1}).
Afterwards, we have computed the impact of each input to the model output by means of the Kernel SHAP algorithm, which is able to recognize the five important features among the twenty defined beforehand; indeed, the descriptors $(x_6, \dots, x_{20})$ have negligible coefficients of importance with respect to the $(x_1, \dots, x_{5})$ ones (Fig.~\ref{friedman2}).

\begin{figure*}
\centering
\includegraphics[width=0.75\textwidth]{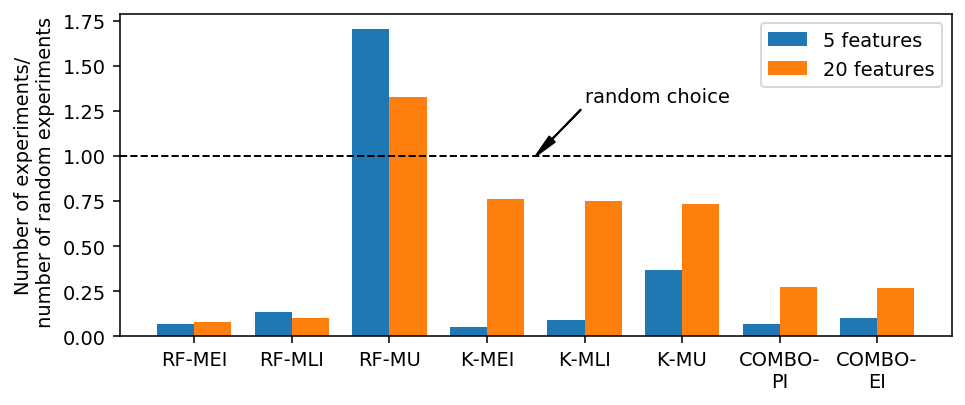}
\caption{Number of evaluations before the convergence to the maximum in the synthetic example for the SL algorithms normalized with respect to the random choice (corresponding to 135 experiments). The initial set consists of the same worst 30 candidates (in terms of the target property) from a random subset of 300 samples of the original dataset.}
\label{friedman3}
\end{figure*}

We have checked the performance of all the regression methodologies (RF-Random Forest, K-Kriging, COMBO) with the related acquisition functions (MEI, MLI, MU, PI, EI) over both the whole synthetic dataset (20 features) and its reduced version (the 5 most important features).
In particular, we aimed at finding the maximum among a random subset of 300 samples from the original 1000 rows long dataset, starting with the pool of 30 points with the lowest target $y$. 
We note that, in general, having a smaller number of descriptors (albeit the meaningful ones) does not allow a faster convergence; anyway, having a feature space with lower dimension implies a lower  computational burden of the SL procedure. 
Figure \ref{friedman3} shows the main results.

\newpage
\subsection*{S2: Lists of the retained variables}
\begin{table}[h]
\centering
\caption{\label{tab2} List of the 44 features retained by AutoMatminer to train the regression model of the Henry coefficient for \ch{CO2}. The reported order is dictated by the SHAP analysis, in accordance with the reported cumulative importance percentage. 
}
\begin{tabular}{lr}
\toprule
  Descriptors & Cumulative\\
  AutoMatminer  &importance\\ 
     &percentage\\
   & (\%) \\
  \midrule
`jml\_log\_vpa' & 19.01\\
`jml\_nn\_75' & 5.56\\
`jml\_first\_ion\_en\_mult\_voro\_coord' & 5.24\\
`jml\_rdf\_22' & 4.35\\
`jml\_C-14' & 3.45\\
`jml\_nn\_15' & 3.23\\
`jml\_nn\_87' & 3.15\\
`jml\_X\_add\_voro\_coord' & 2.96\\
`jml\_rdf\_23' & 2.90\\
`jml\_nn\_25' & 2.83\\
`jml\_nn\_78' & 2.70\\
`jml\_rdf\_76' & 2.61\\
`jml\_nn\_77' & 2.49\\
`jml\_nn\_84' & 2.35\\
`jml\_nn\_79' & 2.33\\
`jml\_nn\_86' & 2.18\\
`jml\_nn\_74' & 2.09\\
`jml\_nn\_82' & 2.07\\
`jml\_rdf\_80' & 2.06\\
`jml\_rdf\_73' & 2.03\\
`jml\_nn\_81' & 1.89\\
`jml\_nn\_76' & 1.88\\
\bottomrule
  \end{tabular}
  \begin{tabular}{lr}
\toprule
  Descriptors & Cumulative\\
  AutoMatminer  &importance\\ 
     &percentage\\
   & (\%) \\
  \midrule
`jml\_nn\_72' & 1.87\\
`jml\_rdf\_54' & 1.80\\
`jml\_X\_mult\_voro\_coord' & 1.59\\
`jml\_rdf\_48' & 1.43\\
`jml\_nn\_85' & 1.32\\
`jml\_nn\_38' & 1.16\\
`jml\_rdf\_61' & 1.07\\
`jml\_rdf\_34' & 1.05\\
`jml\_rdf\_38' & 1.01\\
`jml\_mean\_charge\_13' & 1.00\\
`jml\_rdf\_35' & 0.92\\
`jml\_nn\_80' & 0.83\\
`jml\_nn\_71' & 0.78\\
`jml\_rdf\_60' & 0.76\\
`jml\_nn\_65' & 0.53\\
`jml\_rdf\_21' & 0.52\\
`jml\_pack\_frac' & 0.50\\
`jml\_nn\_21' & 0.42\\
`jml\_rdf\_28' & 0.34\\
`jml\_nn\_50' & 0.32\\
`jml\_nn\_73' & 0.31\\
 `jml\_rdf\_79' & 0.00\\
 
 \bottomrule
  \end{tabular}
\end{table}

\begin{table*}
\centering
\caption{\label{tab4} List of the 92 features retained by AutoMatminer to train the regression model of the working capacity for \ch{CO2}. The reported order is dictated by the SHAP analysis, in accordance with the reported cumulative importance percentage. 
}
\begin{tabular}{lr}
\toprule
  Descriptors & Cumulative\\
  AutoMatminer  &importance\\ 
     &percentage\\
   & (\%) \\
 \midrule
    `jml\_vpa' & 6.80\\
    `jml\_nn\_11' & 4.74\\
    `jml\_nn\_25' & 4.47\\
    `jml\_rdf\_10' & 3.91\\
    `jml\_rdf\_14' & 2.49\\
    `jml\_rdf\_54' & 2.47\\
    `jml\_rdf\_53' & 2.36\\
    `jml\_nn\_22' & 2.13\\
    `jml\_atom\_mass\_divi\_bp' & 2.09\\
    `jml\_nn\_55' & 2.07\\
    `jml\_rdf\_52' & 2.04\\
    `jml\_rdf\_51' & 1.80\\
    `jml\_rdf\_56' & 1.74\\
    `jml\_rdf\_21' & 1.66\\
    `jml\_nn\_38' & 1.57\\
    `jml\_rdf\_24' & 1.49\\
    `jml\_elec\_aff\_mult\_voro\_coord' & 1.49\\
    `jml\_rdf\_58' & 1.44\\
    `jml\_rdf\_15' & 1.37\\
    `jml\_nn\_87' & 1.36\\
    `jml\_rdf\_31' & 1.35\\
    `jml\_nn\_54' & 1.35\\
    `jml\_rdf\_37' & 1.34\\
    `jml\_rdf\_55' & 1.28\\
    `jml\_nn\_40' & 1.24\\
    `jml\_rdf\_60' & 1.23\\
    `jml\_rdf\_59' & 1.21\\
    `jml\_elec\_aff\_mult\_atom\_mass' & 1.20\\
    `jml\_nn\_27' & 1.18\\
    `jml\_nn\_78' & 1.15\\
    `jml\_nn\_33' & 1.11\\
    `jml\_rdf\_57' & 1.04\\
    `jml\_rdf\_13' & 0.94\\
    `jml\_rdf\_69' & 0.94\\
    `jml\_nn\_46' & 0.94\\
    `jml\_rdf\_50' & 0.93\\
    `jml\_rdf\_88' & 0.93\\
    `jml\_rdf\_61' & 0.92\\
    `jml\_nn\_53' & 0.91\\
    `jml\_rdf\_75' & 0.91\\
    `jml\_nn\_29' & 0.90\\
    `jml\_nn\_39' & 0.89\\
    `jml\_nn\_96' & 0.89\\
    `jml\_nn\_92' & 0.88\\
    `jml\_rdf\_78' & 0.87\\
    `jml\_nn\_21' & 0.86\\
  \bottomrule
  \end{tabular}
  \begin{tabular}{lr}
\toprule
  Descriptors & Cumulative\\
  AutoMatminer  &importance\\ 
     &percentage\\
   & (\%) \\
  \midrule
    `jml\_nn\_76' & 0.85\\
    `jml\_rdf\_35' & 0.82\\
    `jml\_rdf\_48' & 0.80\\
    `jml\_nn\_79' & 0.79\\
    `jml\_rdf\_64' & 0.78\\
    `jml\_rdf\_67' & 0.78\\
    `jml\_nn\_75' & 0.76\\
    `jml\_rdf\_40' & 0.75\\
    `jml\_nn\_89' & 0.74\\
    `jml\_nn\_34' & 0.72\\
    `jml\_nn\_30' & 0.67\\
    `jml\_nn\_70' & 0.65\\
    `jml\_nn\_80' & 0.65\\
    `jml\_nn\_77' & 0.61\\
    `jml\_rdf\_39' & 0.54\\
    `jml\_nn\_15' & 0.52\\
    `jml\_rdf\_71' & 0.52\\
    `jml\_hfus\_add\_X' & 0.47\\
    `jml\_rdf\_77' & 0.45\\
    `jml\_rdf\_76' & 0.43\\
    `jml\_rdf\_87' & 0.40\\
    `jml\_elec\_aff\_add\_X' & 0.38\\
    `jml\_rdf\_34' & 0.38\\
    `jml\_nn\_32' & 0.37\\
    `jml\_nn\_73' & 0.37\\
    `jml\_rdf\_36' & 0.37\\
    `jml\_rdf\_86' & 0.36\\
    `jml\_rdf\_73' & 0.36\\
    `jml\_nn\_69' & 0.34\\
    `jml\_nn\_20' & 0.33\\
    `jml\_nn\_66' & 0.32\\
    `jml\_nn\_50' & 0.31\\
    `jml\_mol\_vol\_divi\_atom\_rad' & 0.31\\
    `jml\_nn\_71' & 0.30\\
    `jml\_adf1\_52' & 0.30\\
    `jml\_rdf\_44' & 0.25\\
    `jml\_rdf\_80' & 0.25\\
    `jml\_adf1\_1' & 0.25\\
    `jml\_nn\_72' & 0.24\\
    `jml\_nn\_24' & 0.00\\ 
    `jml\_rdf\_26' & 0.00\\ 
    `jml\_nn\_14' & 0.00\\ 
    `jml\_rdf\_22' & 0.00\\ 
    `jml\_rdf\_23' & 0.00\\ 
    `jml\_rdf\_70' & 0.00\\ 
    `jml\_nn\_16' & 0.00\\ 
\bottomrule
\end{tabular}
\end{table*}

\begin{table}[h]
\centering
\caption{\label{tab3} List of the 36 features retained by AutoMatminer to train the regression model of the Henry coefficient for \ch{H2O}. The reported order is dictated by the SHAP analysis, in accordance with the reported cumulative importance percentage. 
}
\begin{tabular}{lr}
\toprule
  Descriptors & Cumulative\\
  AutoMatminer  &importance\\ 
     &percentage\\
   & (\%) \\
\midrule
`jml\_C-14' & 36.14\\
`jml\_rdf\_22' & 5.41\\
`jml\_rdf\_38' & 5.16\\
`jml\_nn\_10' & 4.92\\
`jml\_nn\_22' & 4.77\\
`jml\_X\_add\_voro\_coord' & 3.58\\
`jml\_log\_vpa' & 3.40\\
`jml\_nn\_13' & 3.06\\
`jml\_nn\_23' & 2.85\\
`jml\_bp\_mult\_atom\_mass' & 2.84\\
`jml\_nn\_20' & 2.63\\
`jml\_nn\_15' & 2.52\\
`jml\_nn\_17' & 2.20\\
`jml\_nn\_16' & 1.86\\
`jml\_rdf\_36' & 1.82\\
`jml\_rdf\_16' & 1.68\\
`jml\_nn\_38' & 1.46\\
`jml\_hfus\_add\_X' & 1.30\\

  \bottomrule
  \end{tabular}
  \begin{tabular}{lr}
\toprule
  Descriptors & Cumulative\\
  AutoMatminer  &importance\\ 
     &percentage\\
   & (\%) \\
  \midrule
  `jml\_nn\_25' & 1.21\\
 `jml\_rdf\_26' & 1.89\\
 `jml\_jv\_enp' & 1.15\\
 `jml\_nn\_79' & 1.10\\
 `jml\_polzbl\_add\_first\_ion\_en' & 1.04\\
 `jml\_nn\_14' & 0.93\\
 `jml\_mol\_vol\_divi\_atom\_rad' & 0.89 \\
 `jml\_adf1\_175' & 0.83\\
 `jml\_nn\_11' & 0.58\\
 `jml\_rdf\_75' & 0.53\\
  `jml\_nn\_81' & 0.50\\
  `jml\_nn\_44' & 0.48\\
  `jml\_nn\_21' & 0.42\\
  `jml\_rdf\_21' & 0.30 \\
  `jml\_rdf\_66' & 0.29\\
  `jml\_nn\_65' & 0.13\\
  `jml\_bp\_divi\_hfus' & 0.00\\
  `jml\_rdf\_80' & 0.00\\
\bottomrule
\end{tabular}
\end{table}

\begin{table*}
\centering
\caption{\label{tab4} List of the 24 features retained by AutoMatminer to train the regression model of the surface area. The reported order is dictated by the SHAP analysis, in accordance with the reported cumulative importance percentage. 
}
\begin{tabular}{lr}
\toprule
  Descriptors & Cumulative\\
  AutoMatminer  &importance\\ 
     &percentage\\
   & (\%) \\
\midrule
    `jml\_vpa' & 46.80\\
    `jml\_pack\_frac' & 11.39\\
    `jml\_nn\_75' & 4.03\\
    `jml\_nn\_81' & 3.95\\
    `jml\_nn\_82' & 3.59\\
    `jml\_nn\_55' & 2.79\\
    `jml\_nn\_67' & 2.54\\
    `jml\_nn\_77' & 2.31\\
  \bottomrule
  \end{tabular}
  \begin{tabular}{lr}
  \toprule
  Descriptors & Cumulative\\
  AutoMatminer  &importance\\ 
     &percentage\\
   & (\%) \\    
 \midrule
    `jml\_nn\_78' & 2.26\\
    `jml\_rdf\_48' & 2.20\\
    `jml\_nn\_80' & 2.17\\
    `jml\_nn\_74' & 1.81\\
    `jml\_nn\_54' & 1.77\\
    `jml\_nn\_72' & 1.69\\
    `jml\_nn\_57' & 1.69\\
    `jml\_nn\_71' & 1.54\\
  \bottomrule
  \end{tabular}
  \begin{tabular}{lr}
  \toprule
  Descriptors & Cumulative\\
  AutoMatminer  &importance\\ 
     &percentage\\
   & (\%) \\    
 \midrule
    `jml\_nn\_73' & 1.43\\
    `jml\_rdf\_53' & 1.09\\
    `jml\_rdf\_73' & 0.84\\
    `jml\_rdf\_45' & 0.83\\
    `jml\_nn\_45' & 0.77\\
    `jml\_rdf\_55' & 0.75\\
    `jml\_rdf\_60' & 0.59\\
    `jml\_rdf\_57' & 0.00\\ 
\bottomrule
\end{tabular}
\end{table*}

\clearpage
\subsection*{S3: List of the retained AutoMatminer operators}

\begin{table}[h]
\centering
\caption{\label{tab:operators} List of the retained AutoMatminer operators. The non tpot.X-type operators are intended to be sklearn operators.}
\begin{tabular}{ll}
\toprule
Feature preprocessors & Feature selectors\\
\midrule
 preprocessing.Binarizer & feature\_selection.SelectPercentile \\
 
 decomposition.FastICA & \\
  cluster.FeatureAgglomeration & \\
  preprocessing.MaxAbsScaler & \\
  preprocessing.MinMaxScaler& \\
  kernel\_approximation.Nystroem & \\
  decomposition.PCA & \\
  preprocessing.PolynomialFeatures & \\
  kernel\_approximation.RBFSampler & \\
  preprocessing.RobustScaler & \\
  preprocessing.StandardScaler & \\
  tpot.builtins.ZeroCount & \\
  tpot.builtins.OneHotEncoder & \\
  \midrule
  ML-models & \\
  \midrule
  ensemble.ExtraTreesRegressor &\\
ensemble.GradientBoostingRegressor &\\
tree.DecisionTreeRegressor &\\
neighbors.KNeighborsRegressor &\\
ensemble.RandomForestRegressor &\\
linear\_model.RidgeCV &\\
 \bottomrule
 \end{tabular}
\end{table}

\clearpage
\subsection*{S4: Determination of the proportionality constant between water uptake and surface area}
\begin{figure}[!hb]
\centering
\includegraphics[width=.45\textwidth]{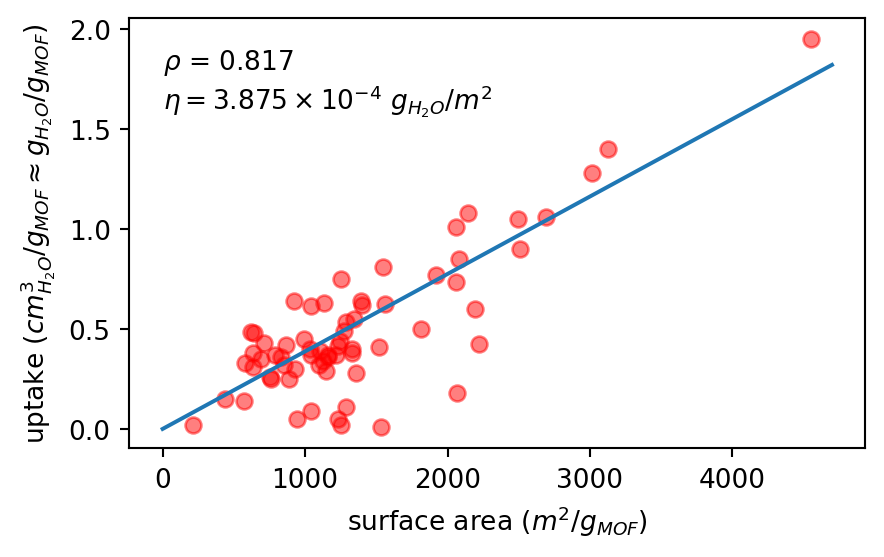}
\caption{Water uptake against surface area for 66 potential MOFs from the database by Chaemchuen \emph{et al.} \cite{chaemchuen2018tunable}. The Pearson coefficient $\rho$ between those two quantities is shown. The linear regression (blue line) gives a constant of proportionality $\eta$, such that $\textrm{uptake}=\eta\times\textrm{surface area}$, in the approximation that $1\, \si{\gram_{\ch{H2O}}}$ occupies a volume of $1\, \si{\cm_{\ch{H2O}}^{3}}$. 
}
\label{fig:corr}
\end{figure}

\ch{H2O} working capacity is necessary to convert the Henry coefficient to units of $\si{\per\pascal}$, but no data are available from the database we referred to.
Therefore we rely on MOFs specific surface area instead.
In detail, Chaemchuen \emph{et al.} have reported \ch{H2O} working capacity for a pool of 66 MOFs \cite{chaemchuen2018tunable}.  
As shown in Fig.~\ref{fig:corr}, a good correlation between the water uptake and the surface area (i.e., the available internal surface per gram of dry adsorbent) can be observed with a Pearson coefficient of $\rho=0.817$, where we have assumed density of \ch{H2O} being $1\, \si{\gram_{\ch{H2O}}\per\cm_{\ch{H2O}}^{3}}$.
On the basis of that correlation, linear regression is imposed for finding the constant of proportionality for the pair water uptake vs the surface area ($\textrm{water uptake}=\eta\times\textrm{surface area}$).
This yields $\eta=3.875\times10^{-4}\, \si{g_{\ch{H2O}}\per\meter\squared}$.

\newpage
\subsection*{S5: Determination of isotherms at different temperatures based on the Polanyi theory}

In the MOFs database we referred to, the Henry coefficient for \ch{H2O} is given only at $T=298\, \si{\kelvin}$, while we needed isotherms at $T_A=308\, \si{\kelvin}$ and $T_C=353\, \si{\kelvin}$. To overcome this issue, we have used the Polanyi theory for rescaling the FFG modeled curve at different temperatures. 

\begin{figure}[h]
    \centering
    \includegraphics[width=0.7\textwidth]{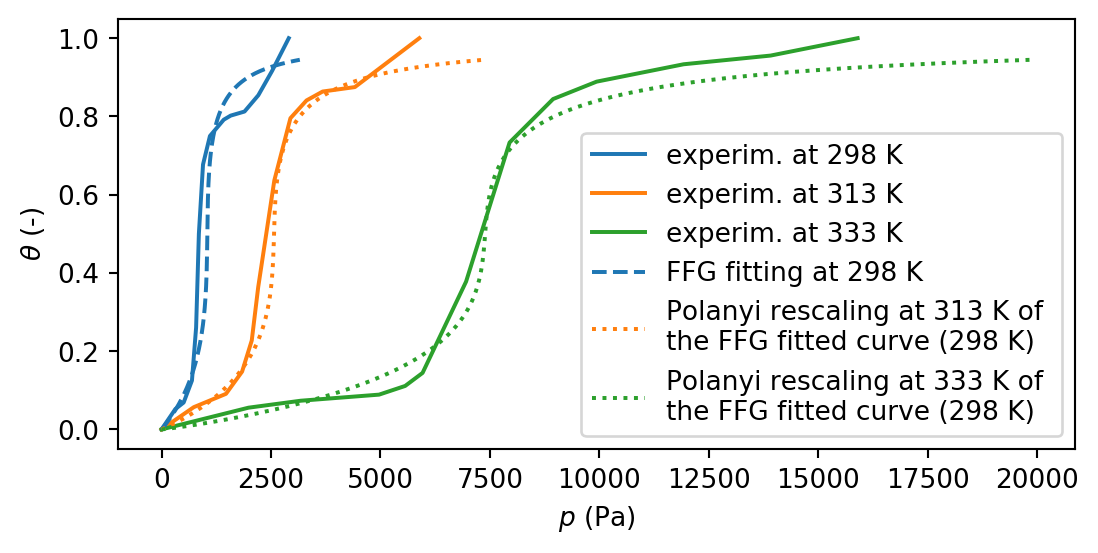}
    \caption{Solid curve: experimental isotherms for water on MOF Al-fumarate at $298\, \si{\kelvin}$ (blue), $313\, \si{\kelvin}$ (orange), $333\, \si{\kelvin}$ (green). Dashed blue curve: FFG fitting of the experimental isotherm at $298\, \si{\kelvin}$ ($H(T=298\, \si{\kelvin}) = 1.4\times10^{-4}\, \si{\per\pascal}$, $\beta = 3.9$). Dotted curves: rescaling of the FFG modeled isotherm (blue dashed curve) at $313\, \si{\kelvin}$ (orange) and $333\, \si{\kelvin}$ (green) according to the Polanyi theory.}
    \label{fig:Polanyi}
\end{figure}

We have checked the validity of this method by means of three experimental isotherms at $298\, \si{\kelvin}$, $313\, \si{\kelvin}$, $333\, \si{\kelvin}$ for the water adsorption on MOF Al-fumarate \cite{jeremias2014advancement}. In particular, we have made a regression over the experimental curve at $298\, \si{\kelvin}$ employing the FFG model with $\beta=3.9$ and $H(T=298\, \si{\kelvin})=1.4\times10^{-4}\, \si{\per\pascal}$ (Fig.~\ref{fig:real_MOFs}, above left). Then, we have rescaled the abscissa of this fitted curve using the Polanyi procedure. Fig.~\ref{fig:Polanyi} shows general agreement between the experimental isotherms and the corresponding rescaled ones at $313\, \si{\kelvin}$ and $333\, \si{\kelvin}$.

\newpage
\newpage
\subsection*{S6: Determination of the average $\beta$ for the FFG model}

\begin{figure*}
\centering
\includegraphics[width=0.39\textwidth]{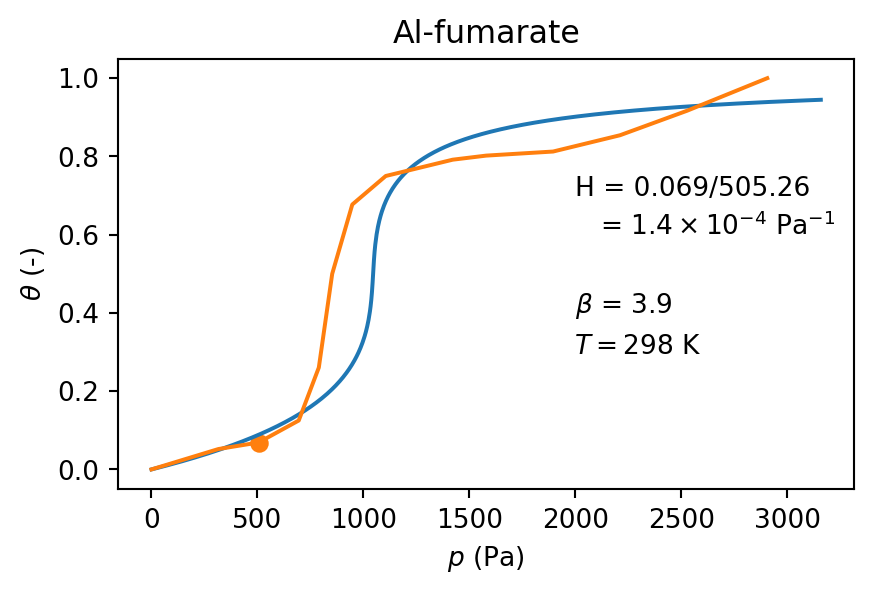}
\includegraphics[width=0.39\textwidth]{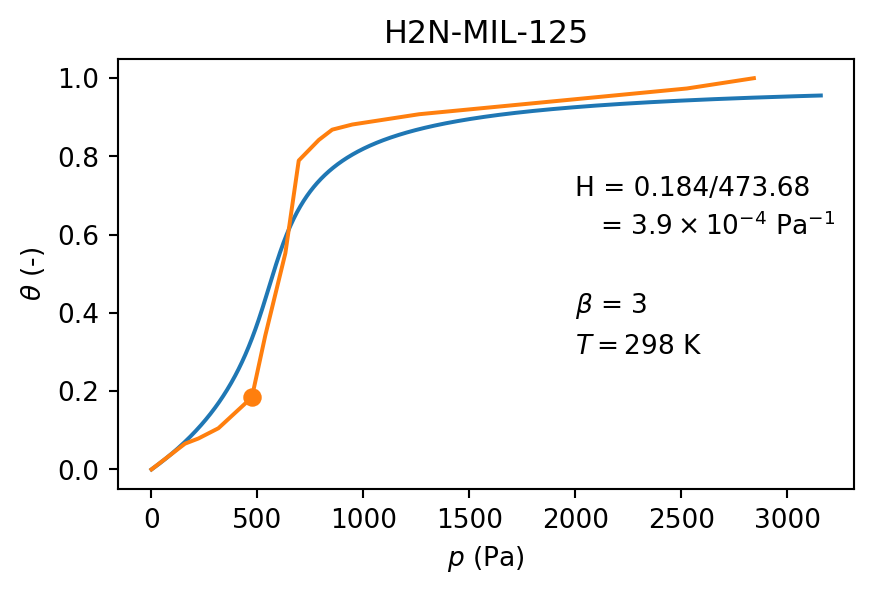}
\includegraphics[width=0.39\textwidth]{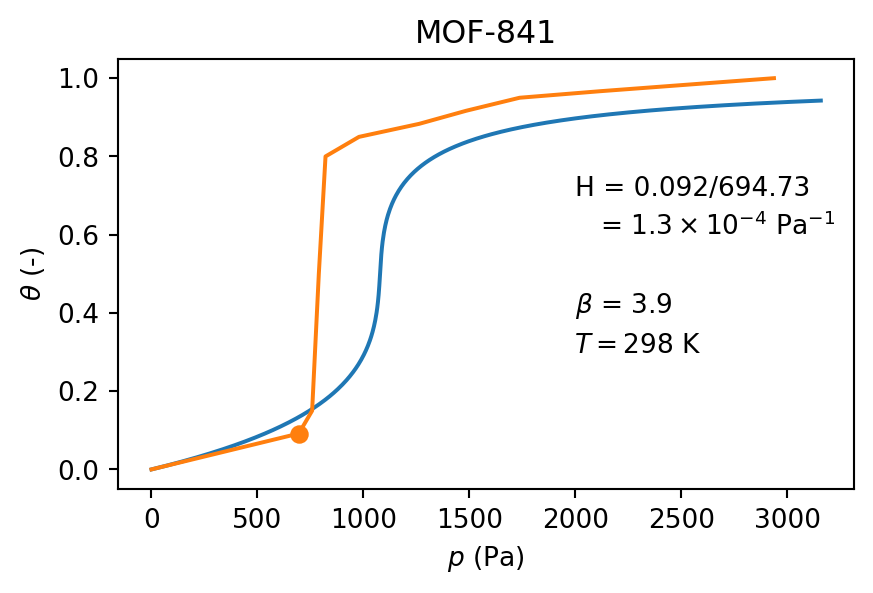}
\includegraphics[width=0.39\textwidth]{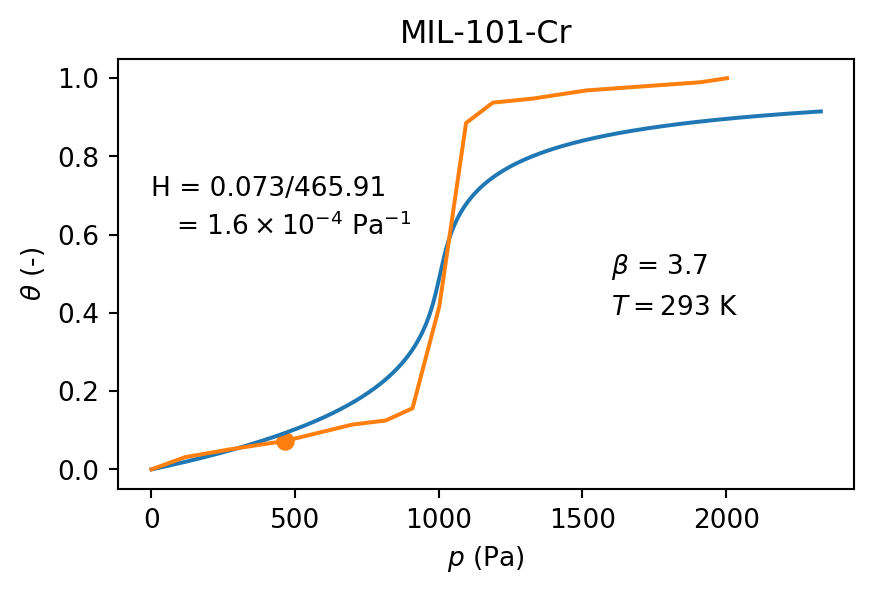}
\includegraphics[width=0.39\textwidth]{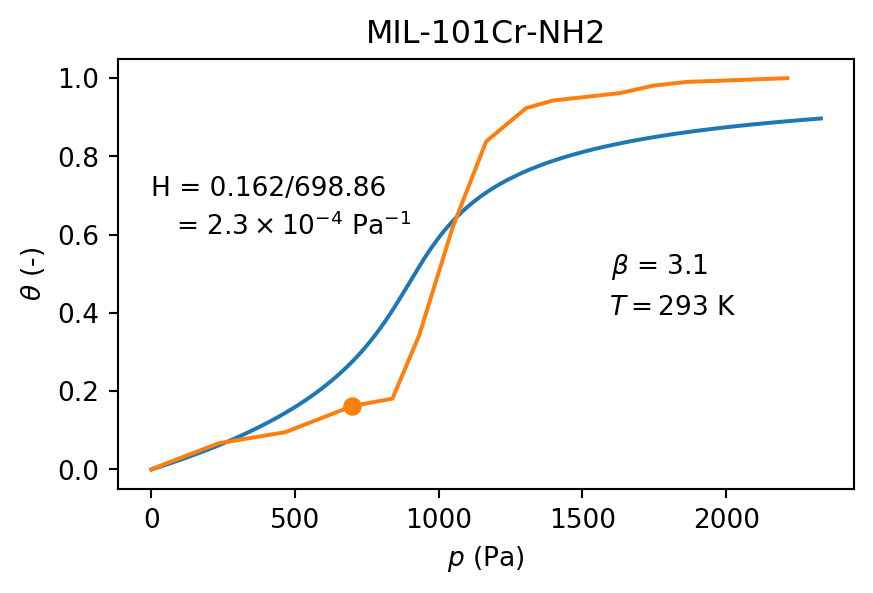}
\includegraphics[width=0.39\textwidth]{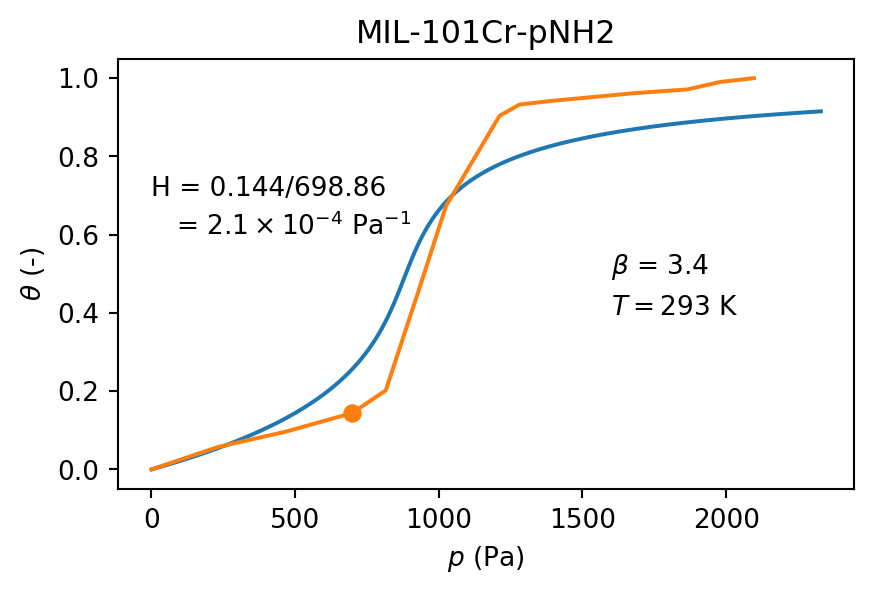}
\includegraphics[width=0.39\textwidth]{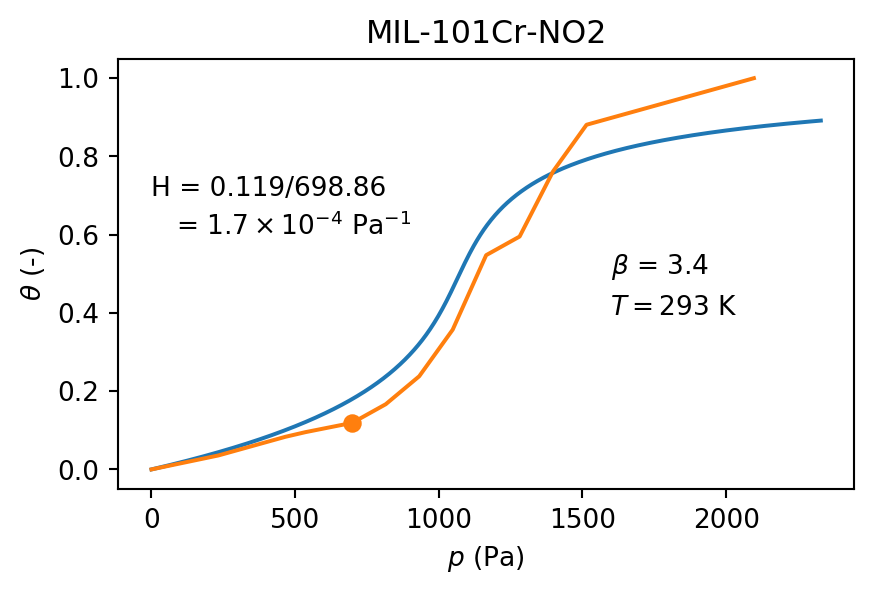}
\includegraphics[width=0.39\textwidth]{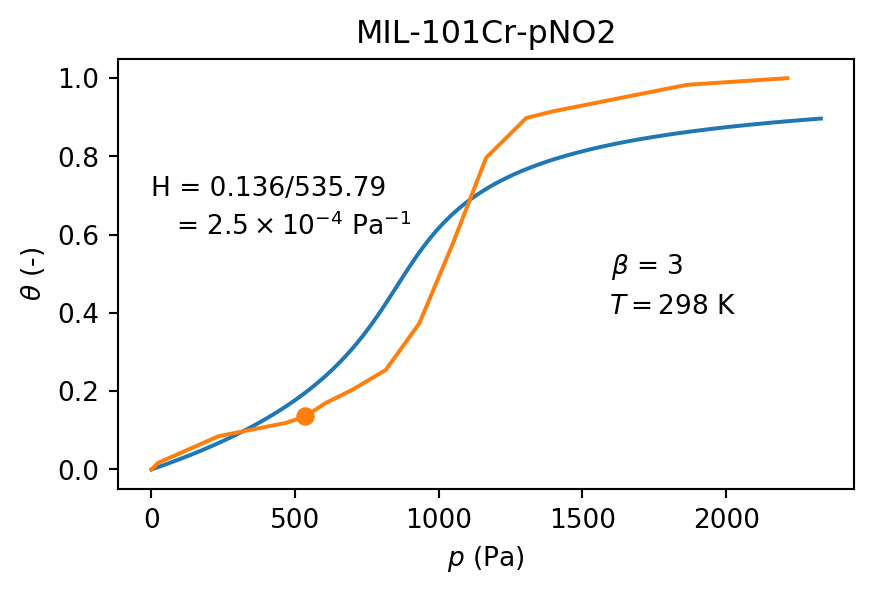}
\caption{Fitting (blue curves) of the experimental isotherms (orange curves) for 8 MOF-water pairs: Al-fumarate \cite{jeremias2014advancement}, \ch{H2N}-MIL-125 \cite{jeremias2013programming}, MOF-841 \cite{furukawa2014water}, MIL-101-Cr, MIL-101Cr-\ch{NH2}, MIL-101Cr-p\ch{NH2}, MIL-101Cr-\ch{NO2}, MIL-101Cr-p\ch{NO2} \cite{khutia2013water}. For each MOF, the orange marker represents the point at which the \ch{H2O} Henry coefficient has been measured as $\theta/p$. The best value of $\beta$ is shown, with the temperature related to the curve.}
\label{fig:real_MOFs}
\end{figure*}

For all the potential MOFs, we have used the average value $\beta=3.4$ in the identification of the isotherm curves according to the FFG model (Eq.~2). For achieving this, we have considered 8 real MOF-water pairs where the isotherm curve (at $298\, \si{\kelvin}$ or $293\, \si{\kelvin}$) is known in the chart ``uptake'' against ``relative pressure $p/p_s(T)$''. We have considered the last experimental point to be at $\theta = 1$, and rescaled the horizontal axis by $p_s(T)$ (Antoine equation), obtaining the usual $\theta - p$ chart. Hence, we have determined, for each of the compounds, the Henry coefficient in units of $\si{\per\pascal}$, by exploiting the low pressure approximation $\theta=H(T)p$. Finally, we have identified for all the MOFs 25 isotherms following the FFG model, with $\beta = 1.5, 1.6, \dots, 3.9$. The best value of $\beta$ has been determined as the one optimizing a least squares problem with respect to the real isotherm. Results are reported in Fig.~\ref{fig:real_MOFs}.

\newpage
\subsection*{S7: Regression methodologies for SL}

\paragraph{Random forest regression.}

In general, a random forest approach \cite{breiman2001random} produces a number of decision trees, consisting in random partitions of the input space. 
On each one of them, a function (i.e., a decision tree) with a low generalization power is built, and their predictions are finally aggregated by taking the mean). The uncertainty of a prediction is a measure of its robustness, i.e., of how much every single known point of the training set influences the predicted value. Here, the Random Forest regression is implemented by means of the Lolo Scala library \cite{hutchinson2016citrine} and does not allow deterministic predictions (i.e., different runs with the same training set give, in general, different responses). \par
\begin{figure*}[hb]
\centering
\includegraphics[width=0.9\textwidth]{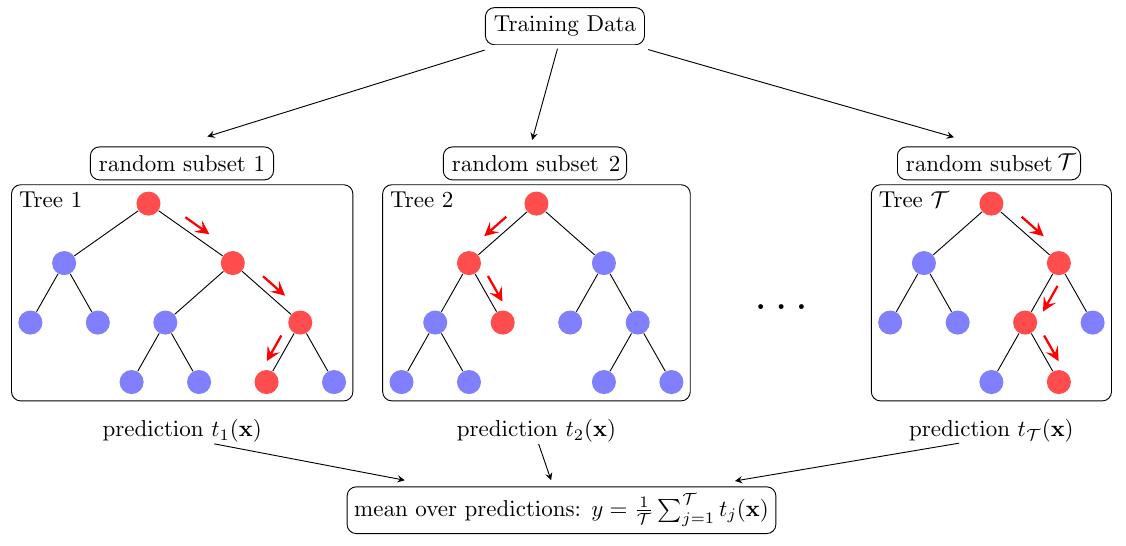}
\caption{Random Forest scheme. Different random subsets of the training data are used to train different decision trees, each one with a low predictive power. Those predictions are aggregated by taking the mean.}
\label{fig:RF-scheme}
\end{figure*}

In particular, Ling \emph{et al.} \cite{hutchinson2016citrine}, following the approach by Efron \cite{efron2012model} and Wager \emph{et al.} \cite{wager2014confidence}, define the uncertainty as
\begin{equation}
\sigma(\mathbf{x}) = \sqrt{\left(\sum_{i=1}^n\max[\sigma_i^2(\mathbf{x}), \omega]\right)+\widetilde{\sigma}^2(\mathbf{x})}
\end{equation}
where $\sigma_i^2(\mathbf{x})$ is the variance at the generic test-point $\mathbf{x}$ due to the $i$-th training point, 
$\omega=|\min_i\sigma^2(\mathbf{x}_i)|$ is a noise threshold and $\widetilde{\sigma}^2(\mathbf{x})$ is an explicit bias function. Specifically:
\begin{equation}
\sigma_i^2(\mathbf{x}) = \textrm{Cov}_j[n_{i,j},t_j(\mathbf{x})]^2+[\overline{t}_{-i}(\mathbf{x})-\overline{t}(\mathbf{x})]^2-\frac{ev}{\mathcal{T}},
\end{equation}
where $\textrm{Cov}_j$ is the covariance over the $j$-th tree, $n_{i,j}$ the number of instances of the $i$-th training point used to fit the $j$-th tree with prediction $t_j(\mathbf{x})$, $\overline{t}_{-i}(\mathbf{x})$ and $\overline{t}(\mathbf{x})$ are the average over trees not trained on sample $i$ and the average over all the trees respectively, $e$ is the Euler's number, $v$ the variance over all the trees, $\mathcal{T}$ the number of trees.  Figure \ref{fig:RF-scheme} shows a scheme of a random forest regression.

\paragraph{Kriging.}
The kriging algorithm, which finds in geostatistics its original field of application \cite{chiavazzo2014reduced}, in the Matlab DACE package implementation \cite{lophaven2002matlab} allows deterministic predictions (i.e., different runs with the same training set give the same response).  
The estimated value of the function is the result of a least squares approach; the uncertainty is the square root of the Mean Squared Error $\sqrt{MSE}$ (as meant by Lophaven \emph{et al.} \cite{lophaven2002matlab}), and depends on the correlation between the stochastic processes modeling the deviations from the predicted value. \par
In particular, a regression model $\mathcal{F}(\bm{\beta}, \mathbf{x})$ is given as a linear combination of $r$ chosen functions
\begin{equation}
\mathcal{F}(\bm{\beta}, \mathbf{x})=\mathbf{g}(\mathbf{x})^T\bm{\beta}=\beta_1g_1(\mathbf{x})+\dots+\beta_{r}g_r(\mathbf{x})
\label{eq1}
\end{equation}
where $\bm{\beta}$ is the array of regression parameters; a suitable stochastic process $z(\cdot)$ is also defined in such a way that, given two domain points $\mathbf{x},\mathbf{w}\in\mathbb{R}^d$, the following holds:
\begin{equation}
\mathbb{E}[z(\mathbf{x})z(\mathbf{w})]\propto\mathcal{R}(\bm{\gamma}, \mathbf{x}, \mathbf{w}),
\end{equation}
where $\mathbb{E}[\cdot]$ denotes the expected value, $\bm{\gamma}$ is the array of parameters governing the correlation model $\mathcal{R}(\bm{\gamma}, \mathbf{x}, \mathbf{w})$; for the typical Gaussian shape (used here), it takes the form $\mathcal{R}(\bm{\gamma}, \mathbf{x}, \mathbf{w})=\prod_{j=1}^{d}\exp(\gamma_j(w_j-x_j)^2)$. \par
For the set of the known $\mathbf{x}_1, \dots, \mathbf{x}_n$, we consider the matrix $\mathbf{G}=[\mathbf{g}(\mathbf{x}_1), \dots, \mathbf{g}(\mathbf{x}_n)]^T$, with $\mathbf{g}(\mathbf{x})$ defined in Eq.~\ref{eq1}. Given the array of known responses $\mathbf{y}=[y_1,\cdots, y_n]^T$ and the corresponding errors $\mathbf{Z}=[z_1, \cdots \mathbf, z_n]^T$, the mean value at the generic point $\mathbf{x}$ is computed by means of the linear predictor $\mu(\mathbf{x})=\mathbf{c}^T\mathbf{y}$, where $\mathbf{c}\in\mathbb{R}^n$ comes from the minimization of the error $\mu(\mathbf{x})-y(\mathbf{x})=\mathbf{c}^T\mathbf{Z}-z(\mathbf{x})+(\mathbf{G}^T\mathbf{c}-\mathbf{g}(\mathbf{x}))^T\bm{\beta}$. The mean squared error (MSE) for the unbiased predictor is strightforwardly $\sigma^2(\mathbf{x})=\mathbb{E}[(\mu(\mathbf{x})-y(\mathbf{x}))^2]=\mathbb{E}[(\mathbf{c}^T\mathbf{Z}-z(\mathbf{x}))^2]$.\par
Here, we have chosen a constant regression model $\mathcal{F}(\bm{\beta}, \mathbf{x})$, with $r=1$, $g_1(\mathbf{x})=1$, the initial guess of the correlation parameter $\bm{\gamma}=\{1\}_d$, with bounds of $10^{-5}$ and $2$. 

\paragraph{Gaussian Process Regression.} Gaussian Processes (GPs) construct a multivariate Gaussian distribution for any point to be evaluated; in this COMmon Bayesian Optimization Library (COMBO, \url{https://github.com/tsudalab/combo3}) implementation is not deterministic. \par
In particular, for the array of the function values $(\mathbf{y}, y_{n+1})^T\in\mathbb{R}^{n+1}$, $\mathbf{y}\in\mathbb{R}^n$ are the known responses, while $y_{n+1}\in\mathbb{R}$ is unknown. A kernel (covariance) matrix is constructed with entries $K_{ij}$ depending on the distances between the arguments $||\mathbf{x}_i-\mathbf{x}_j||$ for the values $y_1,\dots,y_{n+1}$. It gives rise to the zero-centered multivariate normal distribution
\begin{equation}
\begin{pmatrix}
\mathbf{y}\\
y_{n+1}
\end{pmatrix}\sim
\mathcal{N}
\begin{pmatrix}
\mathbf{0},
\begin{pmatrix}
\mathbf{K}+\sigma_y^2\mathbf{I} & \mathbf{k}\\
\mathbf{k}^T & k
\end{pmatrix}
\end{pmatrix}
\end{equation}
where, given the zero-centered Gaussian distributed noise $\epsilon\sim\mathcal{N}(0, \sigma_y^2)$ on the perturbed function $y=f(\mathbf{x})+\epsilon$ and the identity matrix $\mathbf{I}$, the predicted value is $\mu(\mathbf{x}_{n+1})=\mathbb{E}[y_{n+1}]=\mathbf{k}^T(\mathbf{K}+\sigma_y^2\mathbf{I})^{-1}\mathbf{y}$ with a related variance $\sigma^2(\mathbf{x}_{n+1})=k-\mathbf{k}^T(\mathbf{K}+\sigma_y^2\mathbf{I})^{-1}\mathbf{k}$ \cite{rasmussen2006gaussian}.\par
 Here we have used the default squared exponential kernel provided by the COMBO library, by which $K_{ij}=\sigma_0^2\exp(||\mathbf{x}_i-\mathbf{x}_j||^2\ell^{-2}/2)$, where $\sigma_0$ (scale factor) and $\ell$ (width factor) are hyperparameters tuned automatically by means of an embedded stochastic-based ADAM optimization algorithm \cite{ueno2016combo, kingma2014adam}. COMBO, indeed, does not allow deterministic predictions.

\clearpage
\subsection*{S8: Query strategies}
%
\subsubsection*{S8.1 Strategies for Random Forest and Kriging}
Ling \emph{et al.} \cite{ling2017high}, through different combinations of the predicted value $\mu(\mathbf{x}_i)$ and variance $\sigma(\mathbf{x}_i)$, propose three strategies for selecting the next candidate to test: \emph{Maximum Expected Improvement} (MEI), \emph{Maximum Uncertainty} (MU) and \emph{Maximum Likelihood Improvement} (MLI). \par
\begin{itemize}
\item MEI strategy chooses the candidate with the maximum predicted value, i.e., 
\begin{equation}
\mathbf{x}_{MEI}=\arg\max_{i=n+1,\dots,m} \mu(\mathbf{x}_i).
\end{equation}
\item MU strategy chooses the candidate with the maximum uncertainty, i.e.,
\begin{equation}
\mathbf{x}_{MU}=\arg\max_{i=n+1,\dots,m} \sigma(\mathbf{x}_i).
\end{equation}
\item MLI strategy chooses the candidate that the most likely has a higher value than the best previously tested point, i.e.,
\begin{equation}
\mathbf{x}_{MLI}=\arg\max_{i=n+1,\dots,m}-\frac{|\mu(\mathbf{x}_i)-\max_{j=1,\dots,n}y_j|}{\sigma(\mathbf{x}_i)}.
\end{equation}
\end{itemize}
Since variances tend to be high in poorly explored regions, MEI implements a purely \emph{exploitative} strategy, while MU a purely \emph{explorative} strategy. 
In this context, MLI represents a trade-off between the two needs: it takes into account high-performing candidates (exploitation) with sufficient uncertainty (exploration) to have a high likelihood of getting a better test-point.
In this work, we have used the same query strategies for the kriging regression too.

\subsubsection*{S8.2 Strategies for Gaussian Processes}
We have taken into account two of the possible acquisition functions provided by the COMBO library, i.e., Probability of Improvement (PI) and Expected Improvement (EI).\par
\begin{itemize}
\item Kushner \cite{kushner1964new} defines the Probability of Improvement as a criterion for maximizing the quantity
\begin{equation}
PI(\mathbf{x})=P\left(f(\mathbf{x})\geq\max_{i=1, \dots,n}y_i-\xi\right)=\Phi(Z)
\end{equation}
where $Z=\left(\mu(\mathbf{x})-\max_{i=1, \dots, n}y_i\right)/\sigma(\mathbf{x})$, $\Phi(\cdot)$ denotes the CDF of a normal distribution and $\xi$ is a trade-off parameter, whose choice is left to the user. Therefore, according to the PI strategy, the next point to query fulfils
\begin{equation}
\mathbf{x}_{PI}=\arg\max_{i=n+1,\dots,m}PI(\mathbf{x}_i).
\end{equation}
Since we have chosen $\xi=0$, this strategy is purely exploitative.

\item Mo\v{c}kus \emph{et al.} \cite{mockus1978application} define the Improvement as $I(\mathbf{x})=\max\{0, f(\{\mathbf{x}\}_{n+1}^{m})-\max_{i=1,\dots, n}y_i\}$, which is positive when the next query point is greater than the best tested point so far. Its expected value turns out to be
\begin{equation}
\mathbb{E}[I(\mathbf{x})] =
\begin{cases}
\begin{aligned}
\left(\mu(\mathbf{x})-\max_{i=1,\dots, n}y_i\right)\Phi(Z)+\sigma(\mathbf{x})\phi(Z)  ~\hspace{.1cm} &\textrm{if}~ \sigma(\mathbf{x}) > 0\\
0 ~\hspace{.1cm} &\textrm{if} ~\sigma(\mathbf{x}) = 0
\end{aligned}
\end{cases}
\end{equation}
where $Z=(\mu(\mathbf{x})-\max_{i=1, \dots, n}y_i)/\sigma(\mathbf{x})$, while $\phi(\cdot)$ and $\Phi(\cdot)$ denote respectively the PDF and the CDF of a normal distribution. Hence, according to the EI strategy, the next point to query fulfils
\begin{equation}
\mathbf{x}_{EI}=\arg\max_{i=n+1,\dots,m}\mathbb{E}[I(\mathbf{x}_i)],
\end{equation}
which takes into account both the probability of improvement and the potential magnitude of the improvement a point can provide \cite{brochu2010tutorial}.
\end{itemize}

\clearpage

\bibliographystyle{plain}
\bibliography{biblio}